\documentclass[a4paper,12pt, epsfig]{article}
\pdfoutput=1
\usepackage{epsfig}
\usepackage{amssymb,latexsym}
\usepackage{amsfonts}
\usepackage{amsmath}
\usepackage[colorlinks, linkcolor=blue, citecolor=blue, urlcolor=blue]{hyperref}
\usepackage{epstopdf}
\usepackage{graphicx}

\newskip\humongous \humongous=0pt plus 1000pt minus 1000pt

\newif\ifdtup

\catcode`\@=11

\@addtoreset{equation}{section}
\def\theequation{\arabic{section}.\arabic{equation}}

\def\@normalsize{\@setsize\normalsize{15pt}\xiipt\@xiipt
\abovedisplayskip 14pt plus3pt minus3pt%
\belowdisplayskip \abovedisplayskip
\abovedisplayshortskip \z@ plus3pt%
\belowdisplayshortskip 7pt plus3.5pt minus0pt}

\def\small{\@setsize\small{13.6pt}\xipt\@xipt
\abovedisplayskip 13pt plus3pt minus3pt%
\belowdisplayskip \abovedisplayskip
\abovedisplayshortskip \z@ plus3pt%
\belowdisplayshortskip 7pt plus3.5pt minus0pt
\def\@listi{\parsep 4.5pt plus 2pt minus 1pt
      \itemsep \parsep
      \topsep 9pt plus 3pt minus 3pt}}

\relax

\catcode`@=12

\catcode`\@=11

\def\section{\@startsection{section}{1}{\z@}{3.5ex plus 1ex minus
    .2ex}{2.3ex plus .2ex}{\large\bf}}

\def\thesection{\arabic{section}}
\def\thesubsection{\arabic{section}.\arabic{subsection}}

\def\appendix{\setcounter{section}{0}
  \def\thesection{Appendix \Alph{section}}
  \def\thesubsection{\Alph{section}.\arabic{subsection}}
  \def\theequation{\Alph{section}.\arabic{equation}}}

\def\SymBoxes#1#2#3#4{\newdimen\un@t \un@t#3%
\raisebox{#1}{\rule{#2\un@t}{#4}\hskip-#2\un@t
\@tempdimb\un@t \advance\@tempdimb by-#4\@tempcntb#2\relax%
\@whilenum{\@tempcntb>0}\do{
\rule{#4}{\un@t}\hskip\@tempdimb \advance\@tempcntb by\m@ne}%
\hskip-#2\un@t \rule[\un@t]{#2\un@t}{#4}%
\rule[\un@t]{#4}{#4}\hskip-#4
\rule{#4}{\un@t}}\hskip-#4}                

\begin{document}


\newcommand{\dd}{\textrm{d}}

\newcommand{\beq}{\begin{equation}}
\newcommand{\eeq}{\end{equation}}
\newcommand{\bea}{\begin{eqnarray}}
\newcommand{\eea}{\end{eqnarray}}
\newcommand{\beas}{\begin{eqnarray*}}
\newcommand{\eeas}{\end{eqnarray*}}
\newcommand{\defi}{\stackrel{\rm def}{=}}
\newcommand{\non}{\nonumber}
\newcommand{\bquo}{\begin{quote}}
\newcommand{\enqu}{\end{quote}}
\newcommand{\tc}[1]{\textcolor{blue}{#1}}
\renewcommand{\(}{\begin{equation}}
\renewcommand{\)}{\end{equation}}
\def\de{\partial}
\def\Om{\ensuremath{\Omega}}
\def\Tr{ \hbox{\rm Tr}}
\def\rc{ \hbox{$r_{\rm c}$}}
\def\H{ \hbox{\rm H}}
\def\HE{ \hbox{$\rm H^{even}$}}
\def\HO{ \hbox{$\rm H^{odd}$}}
\def\HEO{ \hbox{$\rm H^{even/odd}$}}
\def\HOE{ \hbox{$\rm H^{odd/even}$}}
\def\HHO{ \hbox{$\rm H_H^{odd}$}}
\def\HHEO{ \hbox{$\rm H_H^{even/odd}$}}
\def\HHOE{ \hbox{$\rm H_H^{odd/even}$}}
\def\K{ \hbox{\rm K}}
\def\Im{ \hbox{\rm Im}}
\def\Ker{ \hbox{\rm Ker}}
\def\const{\hbox {\rm const.}}
\def\o{\over}
\def\im{\hbox{\rm Im}}
\def\re{\hbox{\rm Re}}
\def\bra{\langle}\def\ket{\rangle}
\def\Arg{\hbox {\rm Arg}}
\def\exo{\hbox {\rm exp}}
\def\diag{\hbox{\rm diag}}
\def\longvert{{\rule[-2mm]{0.1mm}{7mm}}\,}
\def\a{\alpha}
\def\b{\beta}
\def\e{\epsilon}
\def\l{\lambda}
\def\ol{{\overline{\lambda}}}
\def\ochi{{\overline{\chi}}}
\def\th{\theta}
\def\s{\sigma}
\def\oth{\overline{\theta}}
\def\ad{{\dot{\alpha}}}
\def\bd{{\dot{\beta}}}
\def\oD{\overline{D}}
\def\opsi{\overline{\psi}}
\def\dag{{}^{\dagger}}
\def\tq{{\widetilde q}}
\def\L{{\mathcal{L}}}
\def\p{{}^{\prime}}
\def\W{W}
\def\N{{\cal N}}
\def\hsp{,\hspace{.7cm}}
\def\hspp{,\hspace{.5cm}}
\def\bo{\ensuremath{\hat{b}_1}}
\def\bfo{\ensuremath{\hat{b}_4}}
\def\co{\ensuremath{\hat{c}_1}}
\def\cfo{\ensuremath{\hat{c}_4}}
\def\th#1#2{\ensuremath{\theta_{#1#2}}}
\def\c#1#2{\hbox{\rm cos}(\th#1#2)}
\def\s#1#2{\hbox{\rm sin}(\th#1#2)}
\def\cp#1#2#3{\hbox{\rm cos}^#1(\th#2#3)}
\def\sp#1#2#3{\hbox{\rm sin}^#1(\th#2#3)}
\def\ctp#1#2#3{\hbox{\rm cot}^#1(\th#2#3)}
\def\cpp#1#2#3#4{\hbox{\rm cos}^#1(#2\th#3#4)}
\def\spp#1#2#3#4{\hbox{\rm sin}^#1(#2\th#3#4)}
\def\t#1#2{\hbox{\rm tan}(\th#1#2)}
\def\tp#1#2#3{\hbox{\rm tan}^#1(\th#2#3)}
\def\m#1#2{\ensuremath{\Delta M_{#1#2}^2}}
\def\mn#1#2{\ensuremath{|\Delta M_{#1#2}^2}|}
\def\u#1#2{\ensuremath{{}^{2#1#2}\mathrm{U}}}
\def\pu#1#2{\ensuremath{{}^{2#1#2}\mathrm{Pu}}}
\def\meff{\ensuremath{\Delta M^2_{\rm{eff}}}}
\def\an{\ensuremath{\alpha_n}}
\newcommand{\Z}{\ensuremath{\mathbb Z}}
\newcommand{\R}{\ensuremath{\mathbb R}}
\newcommand{\rp}{\ensuremath{\mathbb {RP}}}
\newcommand{\vac}{\ensuremath{|0\rangle}}
\newcommand{\vact}{\ensuremath{|00\rangle}                    }
\newcommand{\oc}{\ensuremath{\overline{c}}}
\renewcommand{\cos}{\textrm{cos}}
\renewcommand{\sec}{\textrm{sec}}
\renewcommand{\sin}{\textrm{sin}}
\renewcommand{\cot}{\textrm{cot}}
\renewcommand{\tan}{\textrm{tan}}
\renewcommand{\ln}{\textrm{ln}}

\newcommand{\Vol}{\textrm{Vol}}

\newcommand{\half}{\frac{1}{2}}

\def\changed#1{{\bf #1}}

\begin{titlepage}
\bigskip

\def\thefootnote{\fnsymbol{footnote}}

\begin{center}
{\large {\bf
Isolating the Lyman Alpha Forest BAO Anomaly
  } }

\bigskip

\bigskip

{\large \noindent   Jarah Evslin\footnote{\texttt{jarah@impcas.ac.cn}}}

\end{center}

\renewcommand{\thefootnote}{\arabic{footnote}}

\vskip.7cm

\begin{center}
\vspace{0em} {\em  
Institute of Modern Physics, CAS, NanChangLu 509, Lanzhou 730000, China}


\vskip .4cm

\vskip .4cm

\end{center}

\vspace{1.3cm}

\noindent
\begin{center} {\bf Abstract} \end{center}

\noindent
A 2.5-3$\sigma$ discrepancy has been reported between the baryonic acoustic oscillation peak (BAO) in the Lyman $\alpha$ forest at $z\sim 2.34$ and the best fit Planck $\Lambda$CDM cosmology.  To isolate the origin of the tension, we consider unanchored BAO, in which the standard BAO ruler is not calibrated, eliminating any dependence on cosmology before redshift $z\sim 2.34$.   We consider BOSS BAO measurements at $z\sim 0.32$, $0.57$ and $2.34$, using the full 2-dimensional constraints on the angular and line of sight BAO scale, as well as isotropic BAO measurements by 6dF and SDSS at $z\sim 0.106$ and $z\sim 0.15$.  We find that the $z>0.43$ data alone is in 2.9$\sigma$ of tension with $\Lambda$CDM with or without the Planck best fit values of the mass fraction $\Omega_m$ and the BAO scale $r_d H_0$, indicating that the tension arises not from the $\Lambda$CDM parameters but from the dark energy evolution itself at $0.57<z<2.34$.  This conclusion is supported when the acoustic scale measured by the CMB is included, which further increases the tension and excludes a solution with a constant dark energy equation of state.  Including the low $z$ BAO data, which is itself consistent with $\Lambda$CDM, reduces the tension to just over 2$\sigma$, however in this case a CPL parametrization of the dark energy evolution yields only a modest improvement.

\vfill

\begin{flushleft}
{\today}
\end{flushleft}
\end{titlepage}

\hfill{}


\setcounter{footnote}{0}

\section{Introduction} \label{intro}
\noindent

The baryon acoustic oscillation (BAO) peak is an isolated, narrow peak in the two point matter correlation function.  As the matter is nonrelativistic, the peak remains essentially stationary in comoving coordinates \cite{baoteor}.  As a result, it provides a ruler, of fixed comoving length $r_d$ at all redshifts $z$, independent of any cosmological assumptions and of the history of the universe before $z$ and essentially independent of even the gravitational theory, so long as both the matter in question and the light used to observe it are minimally coupled to the metric.  Any appreciable systematic error in the use of this standard ruler must either come from its measurement, via for example gravitational lensing corrections which are known to be small, or from the data analysis, such as the reconstruction technique used to sharpen the peak \cite{eisenrecon,recon2}.

Many analyses of BAO use not only the fact that the ruler's comoving length $r_d$ is independent of $z$, but also use the fact that in the standard cosmological model the ruler's length can be calculated by integrating the speed of sound in the primordial plasma up through recombination and applying a correction for the damping epoch.  Such a calibration of the ruler leads to tighter bounds on cosmological parameters, however it necessarily assumes a detailed cosmological history at high redshift, for example a number of neutrino flavors.  We will refer to a BAO analysis which uses a value of $r_d$ obtained from a cosmological model as {\it{anchored}} BAO.  As a result of this model dependence, anchored BAO is less robust than unanchored BAO, in which one only assumes that $r_d$ is $z$-independent.  Global fits of large quantities of cosmological data generally use anchored BAO.

Recently the BAO feature has been measured in the Lyman $\alpha$ forest of absorption of light from distant quasars, or more precisely in the forest-forest correlation function \cite{ffbao} and in the forest-quasar correlation function \cite{fqbao}.  The effective average redshift of this measurement is $z=2.34$, more than a factor of three greater than any other detection of BAO.  While lower redshift BAO detections have been consistently in excellent agreement with the standard $\Lambda$CDM cosmological model, a discrepancy of $2.5\sigma$ between this measurement and $\Lambda$CDM with the best fit Planck parameters was observed immediately, with as much as $3.5\sigma$ of tension in a global fit \cite{gongboten}.  However, in part because this combination of Planck and BAO used anchored BAO, the analysis of the combined data set used the full cosmological evolution and so there was no clear culprit responsible for this suggestive tension \cite{aubourg}.

There have been analyses of Lyman $\alpha$ forest BAO data using unanchored BAO.  However so far these have either only considered the line of sight Lyman $\alpha$ forest BAO measurement \cite{addison,mebao} or else, in the unanchored analysis, it was assumed that dark energy is in the form of a cosmological constant \cite{aubourg}.  Of these, Ref.~\cite{addison} used older data in which the tension was not yet present while Ref.~\cite{mebao} used too few data points  to break a degeneracy between dark energy and matter density.  The unanchored analysis of Ref.~\cite{aubourg} focused on parameter estimation, rather than an analysis of this tension

In this paper we will answer the following question:  If this anomaly is confirmed by future surveys, just what is its cause?  In other words, the purpose of this paper is to isolate the source of the tension.  To do this, we will use unanchored BAO, and so our results will be sensitive only to the expansion of the universe at $z<2.34$, thus eliminating many of the usual candidate deformations of $\Lambda$CDM such as sterile or massive neutrinos, a running spectral index for primordial perturbations, {\it etc.}  Our results will also be independent of the value of the Hubble parameter.

Most unanchored BAO analyses simply use the ratios of the BAO measurements.  We will instead use the fact that the BAO scale enters only via the parameter
\beq
P=\frac{c}{r_d H_0}
\eeq
which appeared in Ref.~\cite{aubourg} and is equal to $30.0\pm 0.4$ in a best fit Planck $\Lambda$CDM cosmology.  This will allow us to perform an unanchored BAO analysis as follows.  We will fit the full BAO data, not just the ratios.  However we will leave the parameter $P$ free in our fits.  By choosing $P$ to minimize $\chi^2$ (instead of marginalizing), we remove the dependence on the scale $P$, and so our fit of all other parameters will be identical to that obtained using just ratios of BAO measurements.  We prefer this approach to directly fitting the ratios because it yields an additional piece of information, the best fit value of $P$ itself.  We will show that the best fit value of $P$ is in reasonable agreement with $\Lambda$CDM predictions and so the scale is not responsible for the anomaly.  Thus we can conclude for example that the 3$\sigma$ tension between early universe and distance ladder measurements of the Hubble constant \cite{riess} is in fact unrelated to the Lyman $\alpha$ forest BAO anomaly discussed here.

We will find that, using unanchored BAO, there remains $2.9\sigma$ of tension, and so the tension indeed arises from $z<2.34$.  General relativity, or any field redefinition of modified gravity which yields general relativity plus a cosmological fluid, with very weak and standard cosmological assumptions, implies that the only anomalous, isotropic behavior at low redshifts that could affect the observed BAO scale at this epoch has the same effect on the Friedmann equations as dark energy with an equation of state $w(z)\neq -1$, spatial curvature or else a shift in the matter density as a fraction of the critical density $\Omega_m$ or the Hubble constant $H_0$.

\section{The Calculation}

\subsection{BAO Review} \label{revsez}

BAO surveys measure the BAO scale in both the transverse and line of sight directions.  The angular scale of the BAO peak is $\Delta\theta$ while the line of sight scale is determined from the corresponding redshifts, and is called $\Delta z$.  These are related to the location of the peak of the matter two-point function in comoving coordinates $r_d$ via the formulae
\beq
\Delta\theta=\frac{r_d}{(1+z)D_A(z)}\hsp
\Delta z=\frac{r_d H(z)}{(1+z)c} \label{delteq}
\eeq
where $H(z)$ is the Hubble parameter at redshift $z$.  For convenience we decompose the Hubble parameter at redshift $z$ in terms of $H_0$, the Hubble parameter at redshift $z=0$, and a function $E(z)$ as
\beq
H(z)=H_0 E(z)
\eeq
so that the second BAO observable in (\ref{delteq}) becomes
\beq
\Delta z=\left(\frac{r_d H_0}{c}\right) \frac{E(z)}{1+z}= \frac{E(z)}{(1+z)P}
\eeq
where the dimensionless quantity $P=c/(r_dH_0)$ contains all information about the overall length of the ruler in comoving coordinates.

In Eq.~(\ref{delteq}) $D_A(z)$ is the angular diameter distance to redshift $z$ which is given by the integral
\beq
D_A(z)=\left\{\begin{tabular}{ll}
$\frac{c}{(1+z)H_0}\int_0^z\frac{dz^\prime}{E(z^\prime)}$ &if $\Omega_k=0$\nonumber\\
$\frac{c}{(1+z)H_0\sqrt{-\Omega_k}}{\rm{sin}}\left(\sqrt{-\Omega_k}\int_0^z\frac{dz^\prime}{E(z^\prime)}\right)$ &if $\Omega_k\neq 0$\nonumber\\
\end{tabular}\right.
\eeq
so that the first BAO observable in (\ref{delteq}) is
\beq
\Delta\theta=\left\{
\begin{tabular}{ll}
$\frac{1}{P\int_0^z\frac{dz^\prime}{E(z^\prime)}}$ &if $\Omega_k=0$\nonumber\\
$\frac{\sqrt{-\Omega_k}}{P\rm{sin}\left(\sqrt{-\Omega_k}\int_0^z\frac{dz^\prime}{E(z^\prime)}\right)}$ &if $\Omega_k\neq 0$\nonumber\\
\end{tabular}\right.
\eeq
Note that again, all of the dependence on the length of the ruler is contained in the single parameter $P$.  In particular, BAO observations alone cannot disentangle $H_0$ from $r_d$.

The BAO observables are inversely proportional to $P$, and so {\it{ratios}} of BAO scales are independent of $P$, they depend only on the function $E(z)$. The function $E(z)$ on the other hand is determined from Friedmann's equations.  We approximate our Universe to have no radiation and a nonrelativistic matter density which is a fraction $\Omega_m$ of the critical value.  Then, assuming Einstein's theory of gravity and no energy transfer between nonrelativistic matter and other components, the function $E(z)$ is given by
\beq
E(z)=\sqrt{\Omega_m(1+z)^3+\Omega_k (1+z)^2+(1-\Omega_m-\Omega_k)e^{3\int_0^z\frac{1+w(z^\prime)}{1+z^\prime}dz^\prime}}
\eeq
where $w(z)$ is the equation of state of everything except for the nonrelativistic matter, which we will refer to as dark energy.  The parameter $\Omega_k$ is proportional to the spatial curvature today, and again is a fraction of the critical value.

By analogy with the angular diameter distance $D_A(z)$, surveys often define the following distances
\beq
D_H(z)=\frac{c}{H(z)}=\frac{r_d}{(1+z)\Delta z}\hsp
D_V(z)=z^{1/3}(1+z)^{2/3}D_A^{2/3}(z)D_H^{1/3}(z)
\eeq
where $D_V(z)$ is essentially an angle-weighted average of $D_A(z)$ and $D_H(z)$, which is the best determined quantity in measurements of BAO from galaxy surveys.

The surveys then quote the observed ratios
\bea
\frac{D_A(z)}{r_d}&=&\frac{1}{(1+z)\Delta\theta}=
\left\{
\begin{tabular}{ll}
$\frac{P}{1+z}\int_0^z\frac{dz^\prime}{E(z^\prime)}$ &if $\Omega_k=0$\nonumber\\
$\frac{P}{(1+z)\sqrt{-\Omega_k}}{\rm{sin}}\left(\sqrt{-\Omega_k}\int_0^z\frac{dz^\prime}{E(z^\prime)}\right)$ &if $\Omega_k\neq 0$\nonumber\\
\end{tabular}\right.\nonumber\\
\frac{D_H(z)}{r_d}&=&\frac{1}{(1+z)\Delta z}=\frac{P}{E(z)}\nonumber\\
\frac{D_V(z)}{r_d}&=&\frac{1}{1+z}\left(\frac{1+z}{\Delta \theta}\right)^{2/3}\left(\frac{z}{\Delta z}\right)^{1/3}
\eea
Note that all three of these distances are linearly proportional to the constant $P$, and so their ratios depend only upon $E(z)$ or equivalently upon $\Omega_m$, $\Omega_k$ and $w(z)$.  As a result, BAO ratios are particularly robust cosmological tools as they can be used to determine\footnote{There is a degeneracy between these two variables corresponding to the freedom to redefine some of the nonrelativistic matter as dark energy, shifting $\Omega_m$ and/or $\Omega_k$ and changing $w(z)$.  This degeneracy cannot be broken by observations but it has no observable consequences.} $\Omega_m$, $\Omega_k$ and $w(z)$ with no dependence on $H_0$, $r_d$ or the cosmological history at redshifts beyond those measured.

\subsection{Data Sets}

We combine BAO observations from the 6dF Survey \cite{6df}, at an effective redshift of $z=0.106$, with the reconstructed Sloan Digital Sky Survey Data Release 7 \cite{dr7} main galaxy sample (MGS) at an effective redshift of $z=0.15$ and with the Baryon Oscillation Spectroscopic Survey (BOSS) observations in the LOWZ, CMASS and Lyman $\alpha$ forest samples at effective redshifts of $z=0.32$, $z=0.57$ and $z=2.34$ respectively.  The LOWZ and CMASS samples are constructed so as to include galaxies at redshifts below and above $z=0.43$ respectively and so they do not overlap.  On the other hand the MGS sample does have an overlap of several percent with the 6dF and LOWZ samples, but this small overlap is expected to have a negligible contribution to cosmological fits \cite{dr7}.  We have opted not to use data from the WiggleZ survey \cite{wigglez} because of its extensive overlap with BOSS.

At small redshifts, less volume is available and so in particular the MGS BAO determination is already limited by cosmic variance at most of the redshifts considered.  This limitation in statistics implies that these surveys cannot reliably disentangle $D_A(z)$ from $D_H(z)$, and so instead they report only the angle-averaged $D_V(0.106)$ and $D_V(0.15)$.

On the other hand, the three BOSS samples contain sufficient statistics to separate $D_A(z)$ and $D_H(z)$.  They report their results in terms of likelihoods on the $\frac{D_A}{r_d}-\frac{D_H}{r_d}$ plane.  They include $1\sigma$ and $2\sigma$ contours which are roughly elliptical.

For the BOSS CMASS and LOWZ samples, we used the ellipses in the lower two panels of Fig 3 of Ref.~\cite{bosspower}, which were obtained using the BAO feature in the power spectrum. While Ref.~\cite{bosscorr} instead analyzed these datasets using spatial correlation functions, little precision is gained by combining these analyses \cite{bosspower}.  

We determine our $\chi^2$ functions by fitting the 2$\sigma$ ellipses in those figures.  The same exercise with the 1$\sigma$ ellipses yielded essentially the same $\chi^2$ function, showing that the Gaussian approximation is valid at least out to $2\sigma$.  This procedure of course yields the same $\chi^2$ function which may be obtained from the individual uncertainties on $D_A/r_d$ and $D_H/r_d$ together with their covariance.  In Table \ref{datitab} we report this $\chi^2$ function in a slightly unconventional manner: we write the best and least constrained combinations of the radial and angular BAO scales.  These combinations are orthogonal in the coordinates which were used to fit the ellipses.  We believe that this presentation is instructive because we have found that it is the {\it least constrained} data points, corresponding roughly to the Alcock-Pacynski parameters, which drive the preference for dynamical dark energy.


\begin{table}
\centering
\begin{tabular}{|c|l|l|l|l|}
\hline
Data Set&Redshift&Constraint&Ref.&$\Lambda$CDM-Planck\\
\hline\hline
6dF&$z=0.106$&$\frac{D_V(0.106)}{r_d}=2.98\pm 0.13$&\cite{6df}&$3.07\pm 0.04$\\
\hline
MGS&$z=0.15$&$\frac{D_V(0.15)}{r_d}=4.47\pm 0.17$&\cite{dr7}&$4.28\pm 0.06$\\
\hline
BOSS LOWZ&$z=0.32$&$\frac{0.00874D_H(0.32)+0.146D_A(0.32)}{r_d}=1.201\pm 0.021$&\cite{bosspower}&$1.202\pm 0.014$\\
\hline
BOSS LOWZ&$z=0.32$&$\frac{0.0388D_H(0.32)-0.0330D_A(0.32)}{r_d}=0.781\pm 0.053$&\cite{bosspower}&$0.760\pm 0.014$\\
\hline
BOSS CMASS&$z=0.57$&$\frac{0.0158D_H(0.57)+0.101D_A(0.57)}{r_d}=1.276\pm 0.011$&\cite{bosspower}&$1.289\pm 0.014$\\
\hline
BOSS CMASS&$z=0.57$&$\frac{0.0433D_H(0.57)-0.0368D_A(0.57)}{r_d}=0.546\pm 0.026$&\cite{bosspower}&$0.598\pm 0.015$\\
\hline
\end{tabular}
\caption{BAO scale measurements used in this analysis.  All BOSS data uses the most and least precisely determined combination of $D_A$ and $D_H$, chosen to be orthogonal on the $\alpha$ plane in the case of LOWZ and CMASS. The last column lists the $\Lambda$CDM-Planck cosmology predictions.  The Lyman $\alpha$ forest contribution to $\chi^2$ is summarised in Fig.~\ref{chitotfig}.}
\label{datitab}
\end{table}

\begin{figure} 
\begin{center}
\includegraphics[angle=-0,width=2.8in]{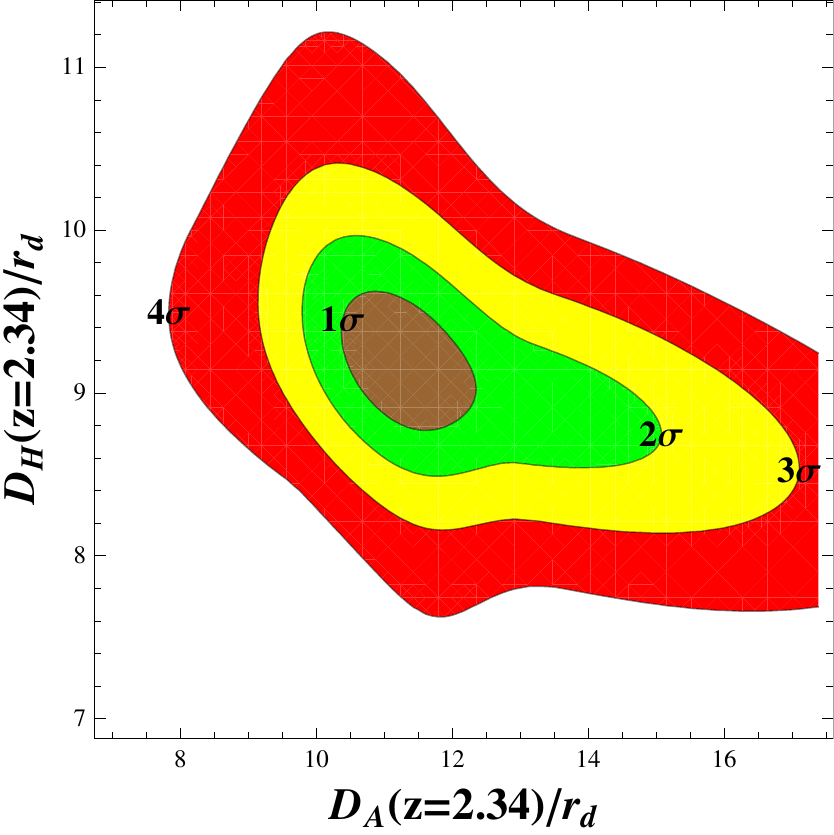}
\includegraphics[angle=-0,width=2.8in]{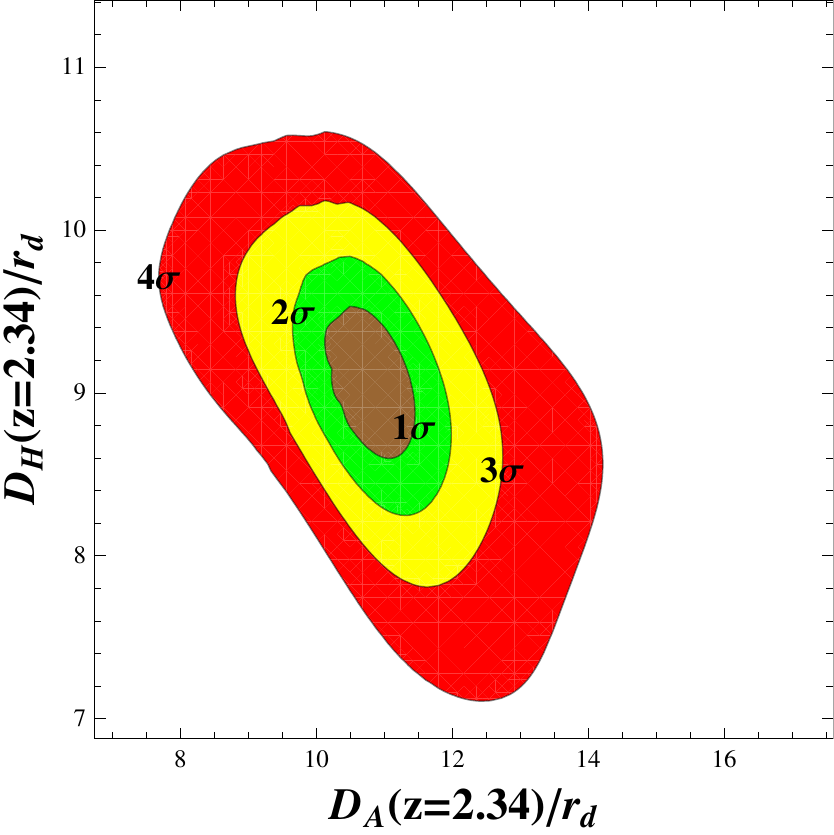}
\caption{The $1\sigma,\ 2\sigma$, $3\sigma$ and $4\sigma$ confidence regions of the line of sight and angular BAO measurements in the (left) Lyman $\alpha$ forest autocorrelation function \cite{ffbao} and (right) Lyman $\alpha$ forest - quasar cross correlation function \cite{fqbao}.}   
\label{lychifig}
\end{center}
\end{figure}

\begin{figure} 
\begin{center}
\includegraphics[angle=-0,width=2.8in]{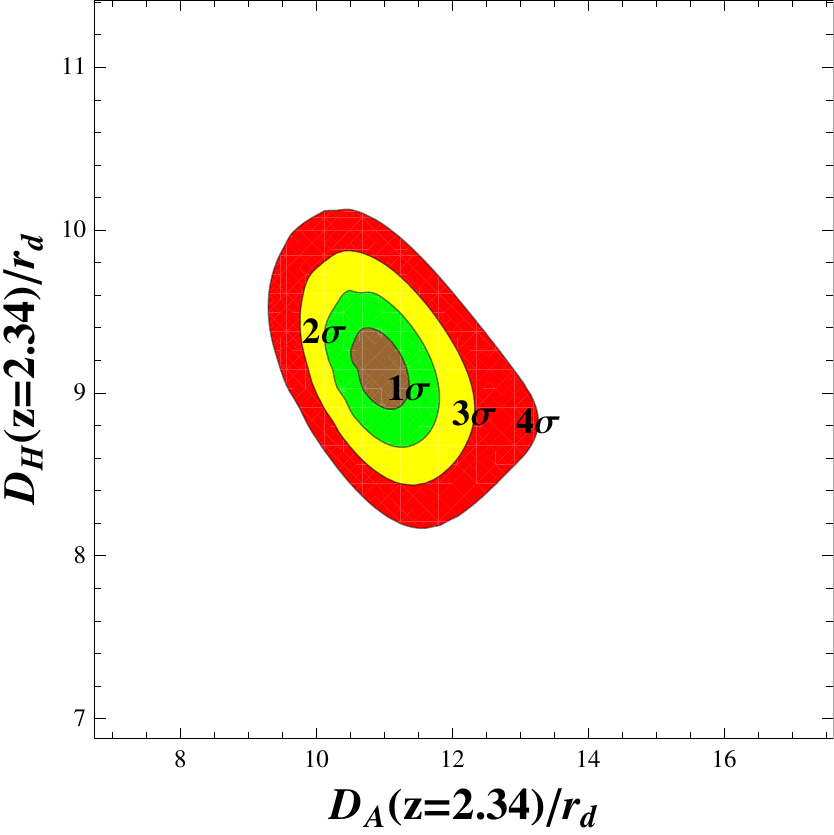}
\caption{The $1\sigma,\ 2\sigma$, $3\sigma$ and $4\sigma$ confidence regions of the line of sight and angular BAO measurements obtained by summing the autocorrelation and cross correlation contributions}   
\label{chitotfig}
\end{center}
\end{figure}

For the Lyman $\alpha$ BAO, in Ref.~\cite{fqbao} the authors measured the BAO scale in the cross-correlation of the Lyman $\alpha$ forest absorption with quasars, at an effective redshift of $z=2.36$, whereas Ref.~\cite{ffbao} measured the BAO scale using the autocorrelation of the Lyman $\alpha$ forest absorption with itself, at an effective redshift of $z=2.34$.  These results are conveniently combined in Fig.~13 of Ref.~\cite{ffbao}.  In part because of the small number of mocks used, the $\chi^2$ function deviates significantly from a quadratic form in these analyses.  Therefore instead of the above quadratic fitting procedure, we have directly used the $\chi^2$ tables supplied on the website~\cite{chitab}.  These are shown in Fig.~\ref{lychifig}.  As these two analyses are quite independent \cite{fqbao} we simply add the $\chi^2$ functions together in our analysis, yielding a total $\chi^2$ shown in Fig.~\ref{chitotfig}.  Note that while the individual contributions exhibit large deviations from the elliptical shape expected in the quadratic approximation, the sum is quite elliptical even at the $3\sigma$ contour.  

Since this note was written, Ref.~\cite{ly13} has appeared with additional data and an improved analysis of the autocorrelation measurement of Ref.~\cite{ffbao}.  They found a 0.5$\sigma$ shift in the line of sight BAO feature which slightly reduces the tension with other measurements.  However the $\chi^2$ data table of that work is not yet public and so we do not use it here.

In summary, our analysis uses eight data points, six of which are summarized in Table~\ref{datitab}.  We use the measurements of $D_V(0.106)/r_d$ and $D_V(0.150)/r_d$ from the 6dF survey and the MGS respectively.  Also we use the determinations of the best and least determined combinations of $D_A/r_d$ and $D_H/r_d$ at $z=0.32$ and $z=0.57$ from BOSS as well as the full $\chi^2$ table for the BOSS Lyman $\alpha$ forest BAO measurement, yielding the $\chi^2$ function on the full $D_H-D_A$ plane.  Our analysis contains more information than most recent BAO analyses, some of which use only $D_V$ at some \cite{planck15,dvnunes} or all redshifts \cite{heavens,dvmuk,dvbas} at which $D_A$ and $D_H$ are available or else use $D_A$ and $D_H$ but not their correlation \cite{dvnunes}.  Some papers \cite{covar} do however use $D_A$ and $D_H$ at each redshift together with their correlation.


On the same table we have also summarized the expected results given a $\Lambda$CDM cosmology with the best fit Planck parameters.  Notice that the tension arises from differences between $D_H$ and $D_A$, corresponding to the parameter combinations which are the least constrained by present data.   The isotropic BAO scale on the other hand is remarkably consistent with the $\Lambda$CDM-Planck predictions. 

\subsection{Methodology}

\begin{table}
\centering
\begin{tabular}{|c|l|l|}
\hline
Cosmology&Parameters&Constraints\\
\hline\hline
$\Lambda$CDM-Planck&None&$\Omega_m=0.312,\ P=30.0,\ w(z)=-1,\ \Omega_k=0$\\
\hline
$\Lambda$CDM&$\Omega_m,\ P$&$w(z)=-1,\ \Omega_k=0$\\
\hline
$w$CDM&$\Omega_m,\ P,\ w$&$w(z)=w,\ \Omega_k=0$\\
\hline
o$\Lambda$CDM&$\Omega_m,\ P,\ k$&$w(z)=-1$\\
\hline
o$w$CDM&$\Omega_m,\ P,\ w,\ k$&$w(z)=w$\\
\hline
CPL&$\Omega_m,\ P,\ w_0,\ w_a$&$w(z)=w_0+\frac{z}{1+z}w_a,\ \Omega_k=0$\\
\hline
\end{tabular}
\caption{Six cosmological models}
\label{modellitab}
\end{table}

We consider the six cosmological models summarized in Table~\ref{modellitab}.  The first is the standard, flat $\Lambda$CDM cosmology with the parameters $\Omega_m$ and $P$ fixed to their best fit values from the Planck experiment \cite{planck15} including lensing and polarization data.  The second again is a flat $\Lambda$CDM model which assumes that dark energy consists of a cosmological constant, so that $w(z)=-1$, but $P$ and $\Omega_m$ are unconstrained.  Of course there are many other free parameters in this model, such as $H_0$, the number of neutrino flavors, the primordial fluctuation spectral parameter, etc.  However, as we have seen in Subsec.~\ref{revsez}, in a homogeneous and isotropic universe, BAO observations only depend on $\Omega_k$, $P$ and $E(z)$ and so on $\Omega_k$, $w(z)$, $\Omega_m$ and $P$.  Third, we consider the same model but allow $w(z)$ to assume any $z$-independent, constant value $w$.  The fourth and fifth models are the same as the second and third except that the curvature parameter $\Omega_k$ is not fixed to zero.  Finally, we consider a flat cosmology with a dark energy equation of state $w(z)$ of the CPL form \cite{cpl1,cpl2}
\beq
w(z)=w_0+\frac{z}{1+z}w_a.
\eeq

Recall that, given our assumptions of homogeneity, isotropy, etc., the measured BAO scale only depends upon $P$, $\Omega_m$, $\Omega_k$ and $w(z)$ therefore the above models are actually somewhat more general than the standard $\Lambda$CDM, $w$CDM, etc.  Any model which satisfies these standard assumptions and has the same $w(z)$ will yield the same results, and so our conclusions may be applied to all such models.  For example, all of the conclusions in this note will apply without alteration in the presence of a sterile neutrino or a second burst of inflation at high redshift.

For each model, we perform two calculations.  First, we determine a goodness of fit of the data to several models.  More precisely, we calculate the $p$-value for a $\chi^2$ statistic defined by
\beq
\chi_j^2=\sum_{i} \frac{(a_{i,j}-b_i)^2}{c_i^2}
\eeq
where $i$ labels the data points from Table~\ref{datitab} and $j$ labels the models from Table~\ref{modellitab}.  The $b_i$ and $c_i$ are the measured values and uncertainties from Table~\ref{datitab}, while each $a_{i,j}$ is the best fit value obtained by choosing the parameters in model $j$ so as to minimize $\chi^2_j$.

If model $j$ is correct, then $\chi_j^2$ can be expected to follow a $\chi^2$ distribution, with a number of degrees of freedom equal to the number of data points used minus the number of parameters in the model $j$, which can be seen in Table~\ref{modellitab}.  This $\chi^2$ distribution is used to calculate the corresponding $p$-value and thus to determine how well model $j$ fits the data.

In addition to significance tests of the six models, we will also perform fits of the parameters in each model, except for the $\Lambda$CDM-Planck cosmology which is defined so as to have no free parameters.  In each case we will choose two free parameters $A$ and $B$ to fit and, if there are other free parameters $C$, they will be chosen so as to minimize $\chi^2$.  We will define $\Delta\chi^2(A,B)$ to be the difference between the value of $\chi^2(A,B)$ with $C$ chosen to minimize $\chi^2$ and the absolute minimum of the $\chi^2$ function for that model.  We will plot the contours on the $A-B$ plane where $\Delta\chi^2(A,B)$ is equal to 2.30, 6.18 and 11.83 which, according to Wilks' theorem, correspond respectively to the boundaries of the $1\sigma$, $2\sigma$ and $3\sigma$ confidence regions for $(A,B)$.

\section{Results}

\subsection{Model Testing}

\begin{table}
\centering
\begin{tabular}{|c|l|l|l|}
\hline
Model&$\chi^2$&Eff. DOF&$p$-value\\
\hline\hline
$\Lambda$CDM-Planck&$15.4$&$4$&$0.0039$\\
\hline
$\Lambda$CDM&$11.1$&$2$&$0.0039$\\
\hline
o$\Lambda$CDM&$11.0$&$1$&$0.0009$\\
\hline
$w$CDM&$3.1$&$1$&$0.078$\\
\hline
\end{tabular}
\caption{$p$ values of $z>0.43$ data with respect to various models.  The effective number of degrees of freedom is the number of data points, 4, minus the number of parameters in the model. The $\Lambda$CDM model is inconsistent with the data at nearly 3$\sigma$ with or without the $\Lambda$CDM-Planck values of the $P$ and $\Omega_m$.  o$\Lambda$CDM is inconsistent at more than 3$\sigma$.}
\label{zaltotab}
\end{table}

\begin{table}
\centering
\begin{tabular}{|c|l|l|l|}
\hline
Model&$\chi^2$&Eff. DOF&$p$-value\\
\hline\hline
$\Lambda$CDM-Planck&$17.3$&$8$&$0.027$\\
\hline
$\Lambda$CDM&$15.0$&$6$&$0.021$\\
\hline
$w$CDM&$8.1$&$5$&$0.15$\\
\hline
o$\Lambda$CDM&$14.9$&$5$&$0.011$\\
\hline
CPL&$7.9$&$4$&$0.094$\\
\hline
\end{tabular}
\caption{$p$ values of all BAO data with respect to various models.  The effective number of degrees of freedom is the number of data points, 8, minus the number of parameters in the model.}  
\label{tuttitab}
\end{table}

We begin by using the BAO data in Table~\ref{datitab} and Fig.~\ref{chitotfig} to test the various modifications of the $\Lambda$CDM standard cosmological model listed in Table~\ref{modellitab}.  As unanchored BAO depends only upon $P$, $\Omega_m$, $\Omega_k$ and $w(z)$, the data in fact only constrains these.  More precisely we will consider two datasets.  First, we will consider a high $z$ data set consisting of the four data points at $z>0.43$, in other words BOSS CMASS and BOSS Lyman $\alpha$.  Next, we will consider all eight data points.

The consistency of the models with the high $z$ data is summarized in Table~\ref{zaltotab}.  As there are four data points, the $p-$value is calculated by comparing the $\chi^2$ statistic to a $\chi^2$ distribution with a number of degrees of freedom equal to four minus the number of parameters in the model.  As the CPL and o$w$CDM models each have four parameters, generically the parameters may be chosen to yield $\chi^2=0$ and correspondingly the number of effective degrees of freedom is equal to zero, therefore these models are not considered.

Our main result can be seen in the first three rows.  $\Lambda$CDM yields a $p$-value of only $0.0039$, meaning that it is inconsistent with the high $z$ data at 2.9$\sigma$.  Imposing the best fit $\Lambda$CDM-Planck values of $\Omega_m$ and $P$ slightly increases $\chi^2$ but doubles the effective number of degrees of freedom, resulting in essentially the same inconsistency.  Adding curvature increases the tension yet further.  This means that the values of $\Omega_m$, $\Omega_k$ and $P$ are not responsible for the tension, instead: {\it The tension reflects the incompatibility of the BAO data with a cosmological constant ($w(z)=-1$) model of dark energy in the range $0.57<z<2.34$.}   In the rest of this note, we will see that our other results are consistent with this conclusion.  Indeed in the last row one sees that by allowing dark energy to assume a constant but arbitrary value of $w$, the tension is already reduced to a statistically less $1.8\sigma$.  

We next consider all eight BAO data points.  The compatibility of the full BAO data with the models is summarized in Table~\ref{tuttitab}.  In the first two rows, one can see that the significance of the anomaly has dropped to $2.2\sigma$ and $2.3\sigma$ respectively.  This is no surprise, as the low $z$ data is known to be quite consistent with both $\Lambda$CDM and with the best fit $\Lambda$CDM-Planck results.  The fact that the significance of the anomaly is reduced when including low $z$ may lead one to believe that the significance has been overstated in the previous analyses, due to the look elsewhere effect \cite{lee}.  Of course the inclusion of data from a consistent regime does reduce the statistical significance of any anomaly.  However the fact that the high $z$ data set in fact covers a large volume and a much larger redshift range than the low $z$ data encourages one to hope that this $2.9\sigma$ anomaly is not just pi in the sky \cite{pi}.

While $2.3\sigma$ is only a slight tension between the data and $\Lambda$CDM, and $2.5\sigma$ is only mild tension with respect to o$\Lambda$CDM, nonetheless the model consistency with the data improves as the dark energy equation of state is unconstrained.  By allowing it to be an arbitrary constant, $w$, one finds only $1.4\sigma$.  
Relaxing the equation of state yet further to the CPL model of course reduces $\chi^2$, but due to the extra degree of freedom it actually yields a slightly worse fit, at $1.7\sigma$.  
In conclusion, while releasing the dark energy equation of state does somewhat reduce the tension, there is no clear preference for the CPL form.  On the contrary, a better fit may be obtained from a model with an abrupt transition between $z=0.57$ and $z=2.34$, as in Ref.~\cite{jaber}, or else with negative dark energy at high redshift, as in Refs.~\cite{ffbao,1405.5116}.

\subsection{Parameter Fitting}

\begin{figure} 
\begin{center}
\includegraphics[angle=-0,width=2.8in]{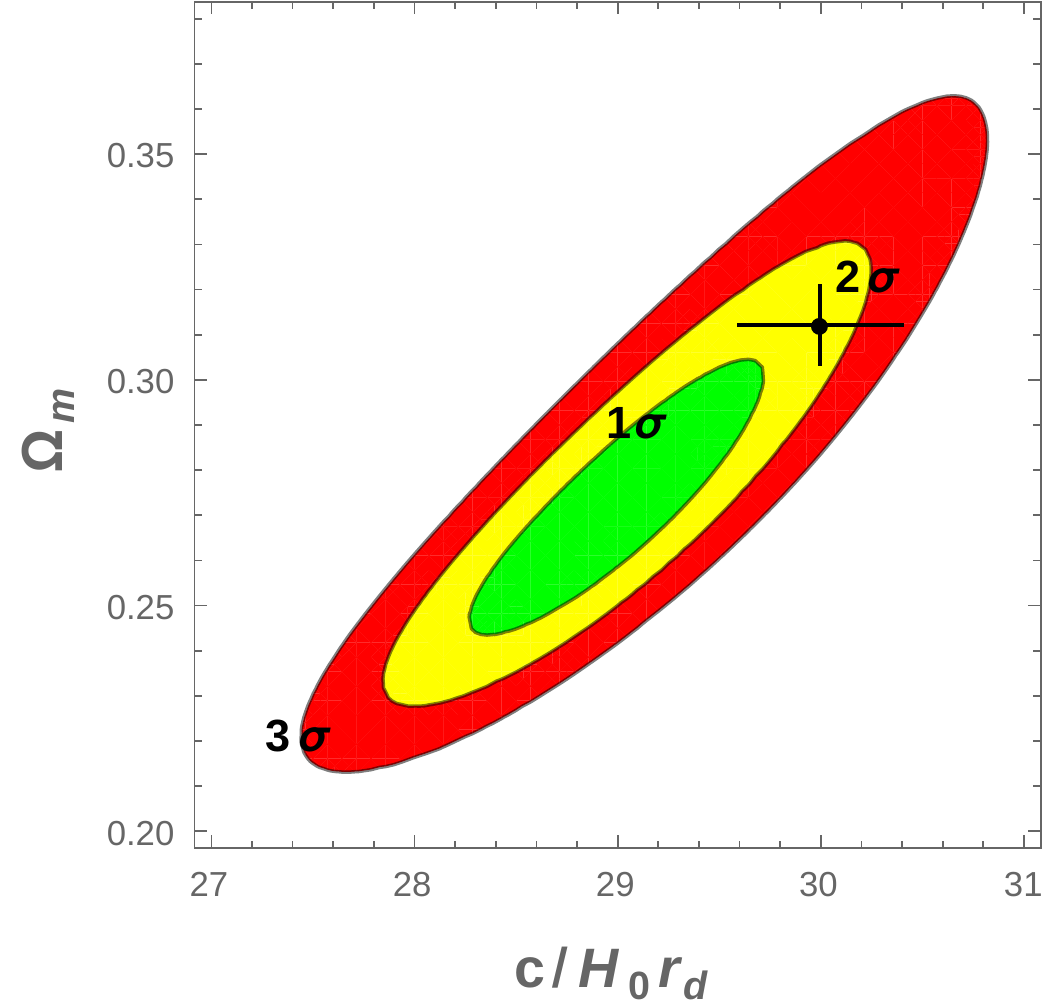}
\includegraphics[angle=-0,width=2.8in]{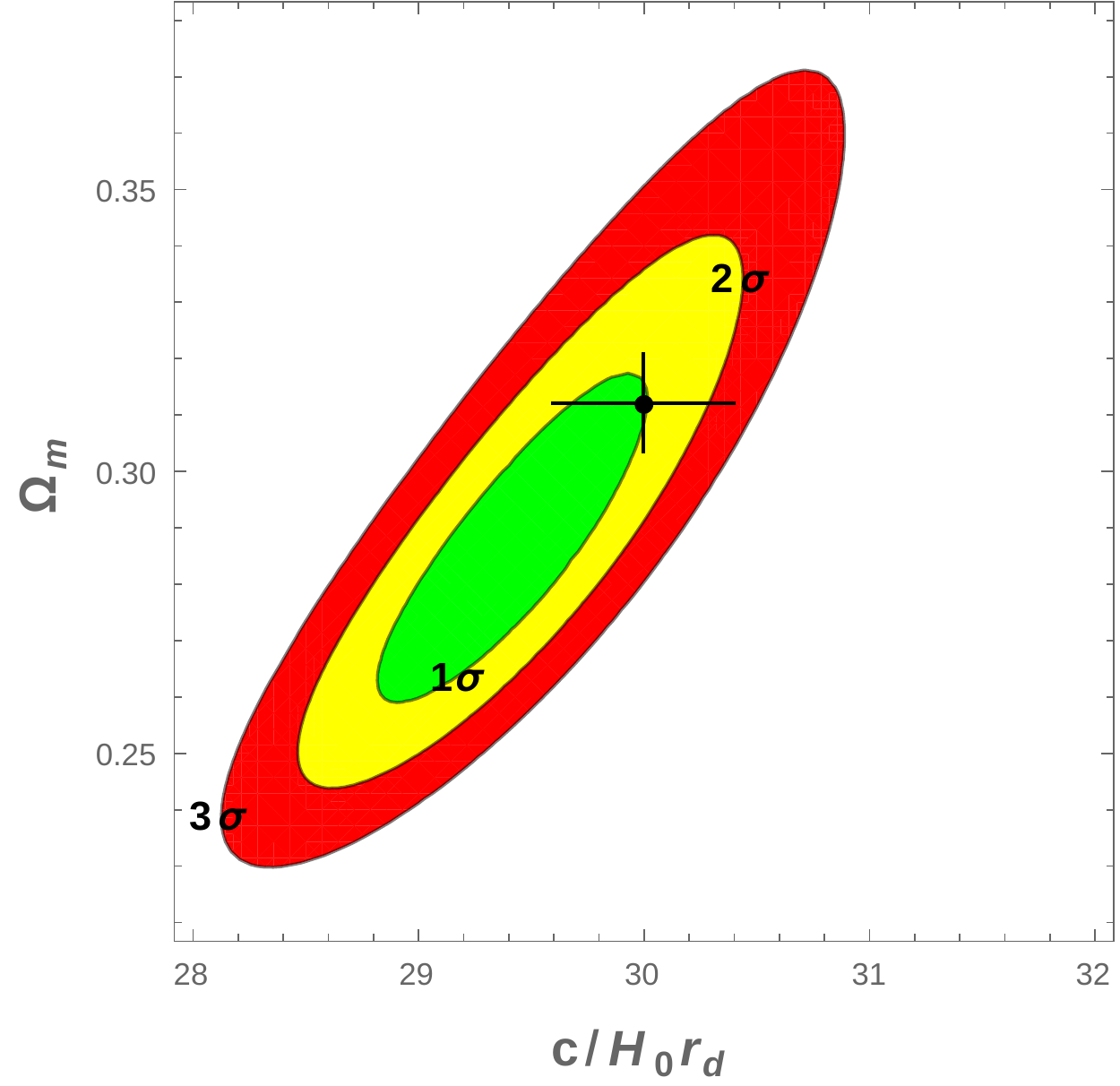}
\caption{The $1\sigma,\ 2\sigma$ and $3\sigma$ confidence regions for a fit of $P=c/(H_0r_d)$ and $\Omega_m$ to the $z>0.43$ data (left) and to all BAO data (right), assuming the $\Lambda$CDM model.  The dot is the $\Lambda$CDM-Planck best fit value with its associated uncertainties.}   
\label{lcdmfig}
\end{center}
\end{figure}

The same BAO data allows one to fit the parameters in each model.  Given the assumptions used in this note, $P$, $\Omega_k$ and $\Omega_m$ are the only parameters that may affect the BAO scale in the $\Lambda$CDM model, or more generally whenever dark energy is a cosmological constant $w(z)=-1$.  In Fig.~\ref{lcdmfig} we plot the $1\sigma$, $2\sigma$ and $3\sigma$ confidence regions for the parameters $P$ and $\Omega_m$ in the $\Lambda$CDM model, given the high $z$ and the full BAO data sets.  We note that the $\Lambda$CDM-Planck best fit values are consistent, at about the $1\sigma$ level, with the BAO best fit values.  This again illustrates that the tension between BAO data and the standard $\Lambda$CDM-Planck cosmology does not arise from the parameters $P$ and $\Omega_m$, it can only arise from $w(z)$.

\begin{figure} 
\begin{center}
\includegraphics[angle=-0,width=2.8in]{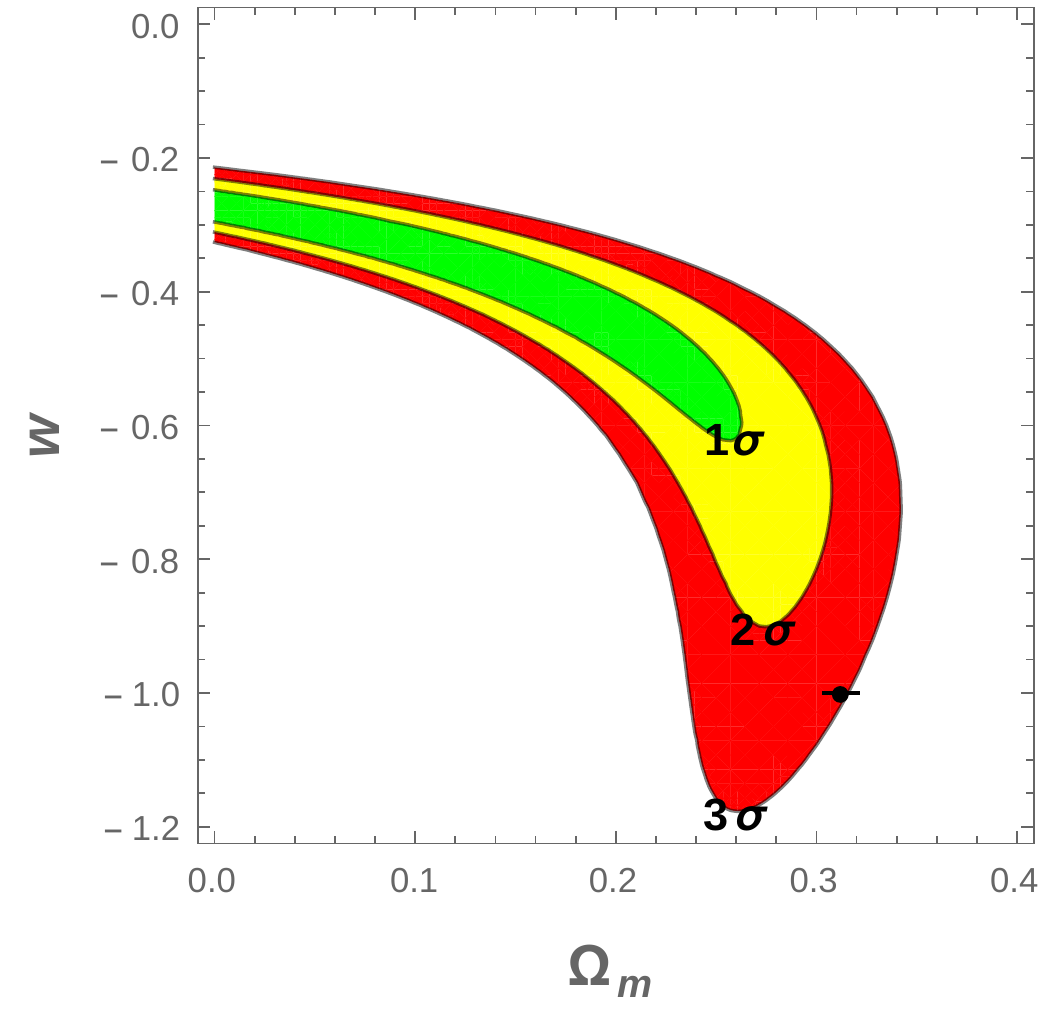}
\includegraphics[angle=-0,width=2.8in]{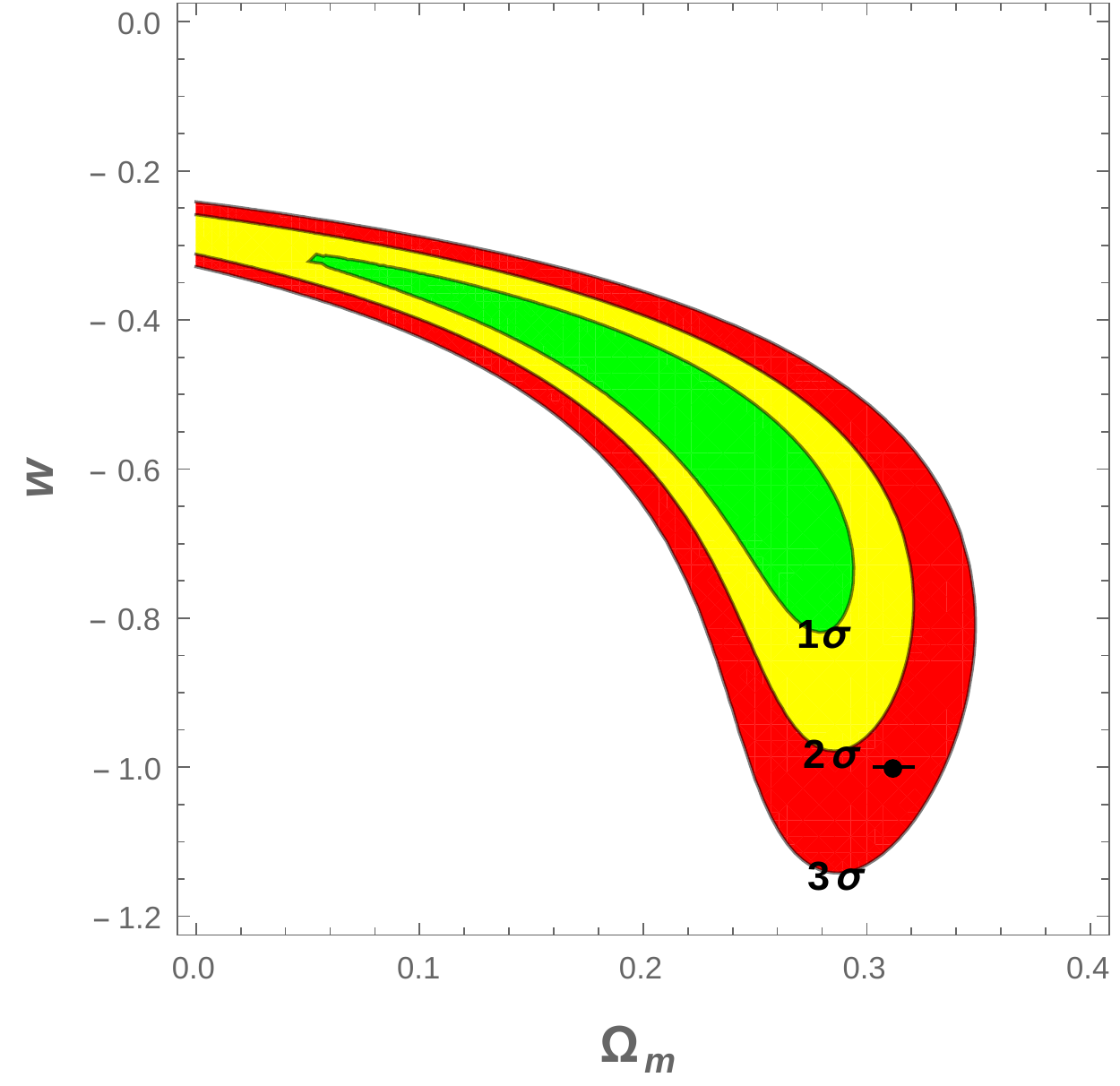}
\caption{The $1\sigma,\ 2\sigma$ and $3\sigma$ confidence regions for a fit of $w$ and $\Omega_m$ to the $z>0.43$ data (left) and to all BAO data (right), assuming the $w$CDM model.  The dot is the $\Lambda$CDM-Planck best fit value with its associated uncertainties.  The $3\sigma$ tension between the $\Lambda$CDM-Planck value and the high $z$ BAO data is our main result.}   
\label{wfig}
\end{center}
\end{figure}

Next in Fig.~\ref{wfig} we find the confidence regions for $\Omega_m$ and $w$ in the $w$CDM model.  At each point, $P$ is chosen to minimize $\chi^2$.  This procedure is equivalent to only considering ratios of BAO data points, which would yield one less degree of freedom and one less parameter $(P)$ and so the same number of effective degrees of freedom.  The most obvious feature of this figure is the good fit provided at low $\Omega_m$.  Of course very low $\Omega_m$ is excluded by a number of cosmological probes, and so the low $\Omega_m$ tail of these plots will not be our focus.  Including the low $\Omega_m$ part of the plots or not, one observes a rather poor agreement with the $\Lambda$CDM-Planck values.  In particular the high $z$ BAO data are in about $3\sigma$ of tension.  The $w$CDM model reduces the tension only by choosing a much higher value of $w$.  A higher (close to zero) value of $w$, so that dark energy is nearly nonrelativistic matter, is favored strongly by the high $z$ data but is somewhat constrained by the low $z$ data.  Nonetheless, one can see that the BAO data clearly prefer a higher value of the dark energy equation of state $w$.

\begin{figure} 
\begin{center}
\includegraphics[angle=-0,width=2.8in]{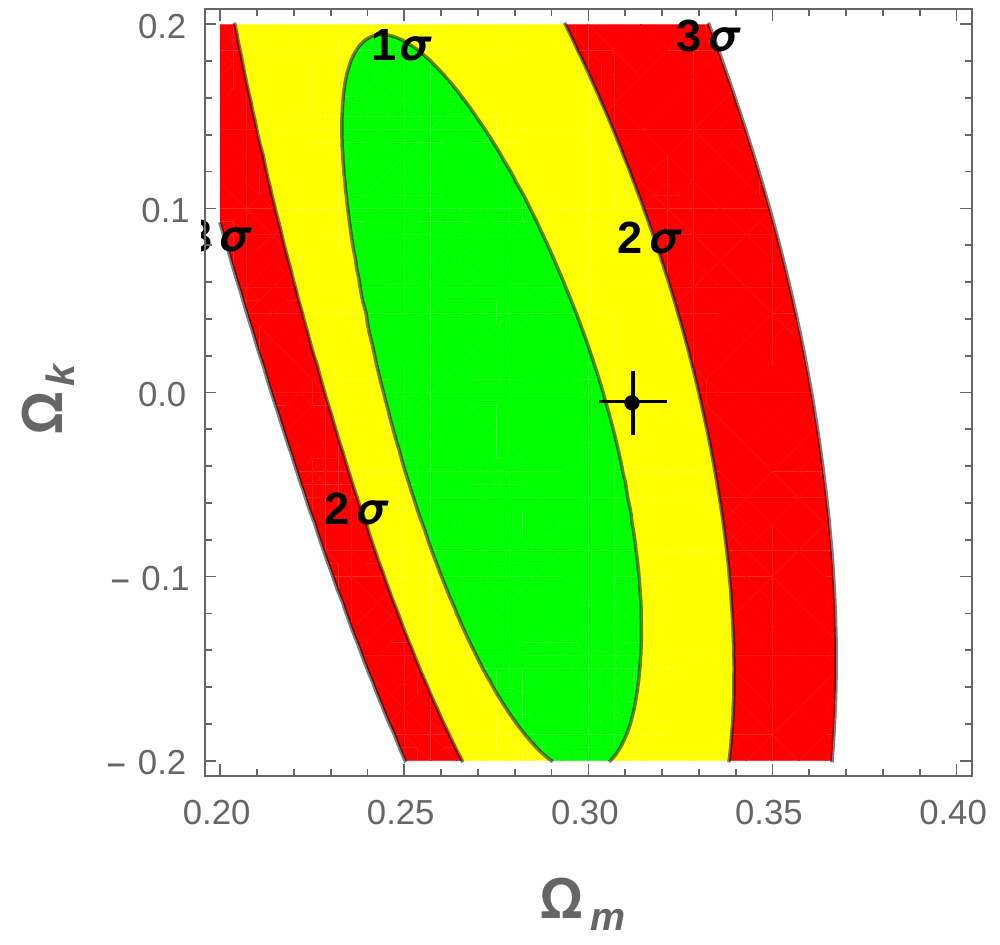}
\includegraphics[angle=-0,width=2.8in]{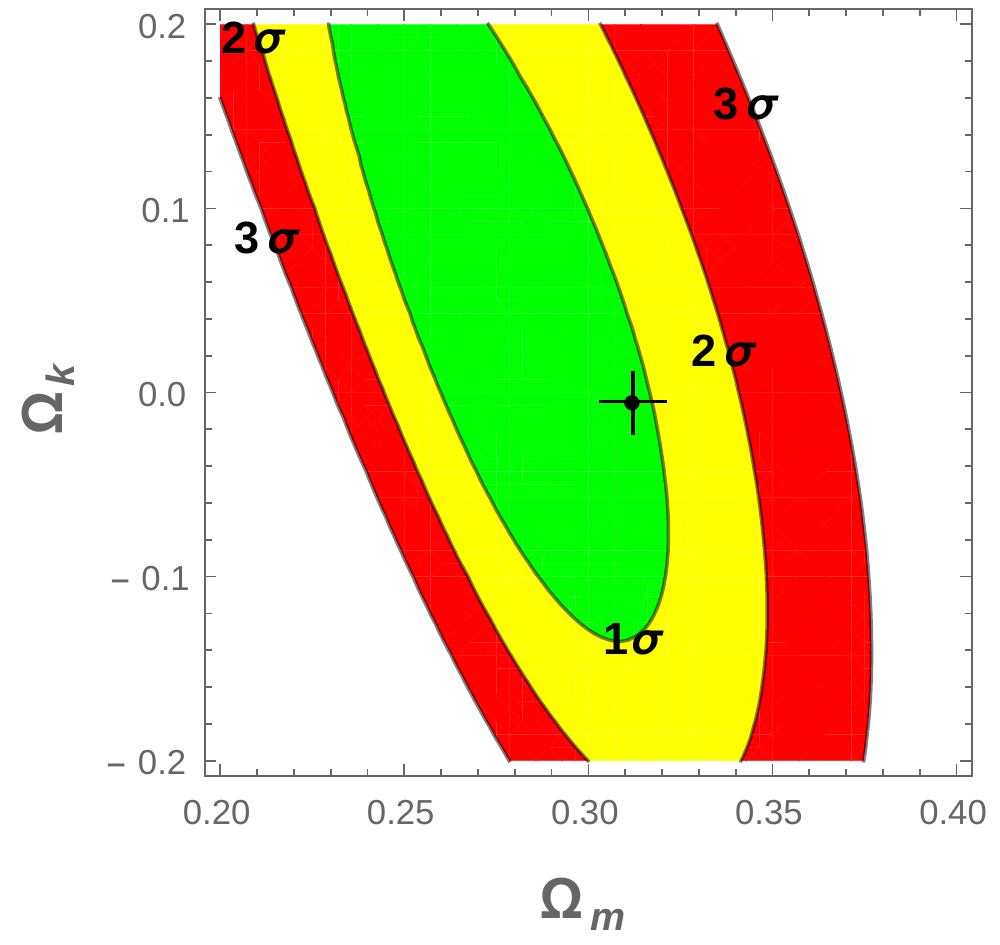}
\caption{The $1\sigma,\ 2\sigma$ and $3\sigma$ confidence regions for a fit of $\Omega_m$ and $\Omega_k$ to the $z>0.43$ data (left) and to all BAO data (right), assuming the o$\Lambda$CDM model.  The dot is the $\Lambda$CDM-Planck best fit value with its associated uncertainty shown for $\Omega_m$ and the o$\Lambda$CDM uncertainty, as reported by Planck \cite{planck15}, shown for $\Omega_k$.}   
\label{kfig}
\end{center}
\end{figure}

On the contrary, in Fig.~\ref{kfig} one sees that the BAO data has only a slight preference for a higher value of the spatial curvature in the o$\Lambda$CDM model.  The smallness of this preference is consistent with the fact that the $p$-values of the $\Lambda$CDM and o$\Lambda$CDM models are similar, and so a modification of curvature without dynamical dark energy does not ease the tension with the BAO data.  

\begin{figure} 
\begin{center}
\includegraphics[angle=-0,width=2.8in]{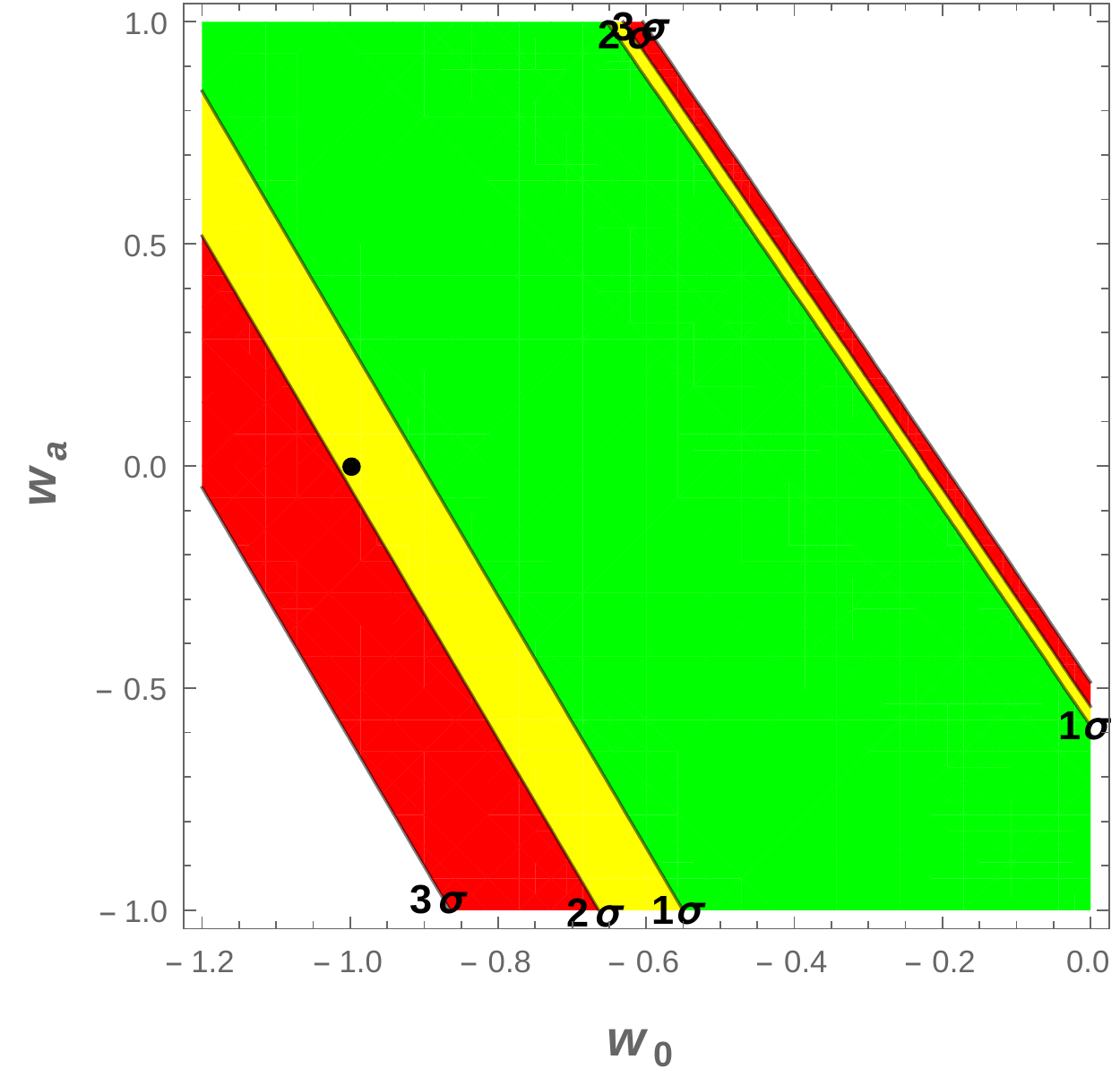}
\caption{The $1\sigma,\ 2\sigma$ and $3\sigma$ confidence regions for a fit of $w_0$ and $w_a$ to all of the BAO data, assuming the CPL model.}
\label{cplfig}
\end{center}
\end{figure}

Finally in Fig.~\ref{cplfig} we confront the CPL dynamical dark energy parametrization with the full BAO data.   At each point the parameters $\Omega_m$ and $P$ are chosen so as to minimize $\chi^2$.  $P$ depends only upon the overall BAO scale whereas the other parameters depend on the ratios of BAO measurements.  Therefore the choice of $P$ does not affect the fitting of the other parameters.  The situation is somewhat more complicated for $\Omega_m$ due to a near degeneracy.  As a result, in a Bayesian approach our results would be strongly dependent upon the prior assigned to $\Omega_m$.   

We can see about 2$\sigma$ of tension with a cosmological constant.  There is a broad degeneracy band with a sharp cutoff at the upper-right.  This sharp cutoff results from the fact that, as one moves up and right, the best fit value of $\Omega_m$ is reduced, but we do not allow it to be less than zero.  The sharp cutoff occurs when the best fit $\Omega_m$ reaches zero.

While it appears  as though there is a broad degeneracy which allows little constraint on the CPL parameters, in fact most of this allowed region has a best fit value of $\Omega_m$ so low that it is in contradiction not only with $\Lambda$CDM-Planck, but even with measurements of dark matter in galaxy clusters.  Fixing $\Omega_m$ to its $\Lambda$CDM-Planck value, which provides a more realistic parameter region, the compatibility with a cosmological constant remains about $2\sigma$.  In this case there is a clear preference for a higher current value of the dark energy equation of state with a lower value at high redshift, consistent with the results of previous analyses \cite{mebao}.

Overall, the CPL parametrization yields only a modest reduction in the tension.  A better fit may arise from a more exotic parametrization such as that of Refs.~\cite{greci,jaber}, however as the anomaly arises from $2.34>z>0.57$, to distinguish such alternatives one must await precise BAO measurements in this regime.  Fortunately one need not wait long \cite{gongboeboss}.

\section{Cosmic Microwave Background}

\subsection{Consistency and Higher Redshifts}

Is any set of observed values of $\Delta\theta$ and $\Delta z$ consistent with some cosmology?  No.  General relativity coupled to a suitably chosen matter sector can produce any evolution $\rho(z)$ and $p(z)$ of the density and pressure and so any evolution of the Hubble parameter $H(z)$.  This means that any evolution of $D_A(z)$ {\it or} $D_H(z)$ is possible.  However, since only one function $H(z)$ is free, and it entirely determines $D_A(z)$ and $D_H(z)$ up to some constants, there must be one constraint relating $D_A(z)$ and $D_H(z)$, and so $\Delta\theta$ and $\Delta z$, so that the number of degrees of freedom is equal to unity.

This constraint is not difficult to derive from the formulae in Subsec.~\ref{revsez}.  Writing
\beq
(1+z)D_A(z)=\frac{c}{H_0\sqrt{-\Omega_k}}{\rm{sin}}\left(H_0\sqrt{-\Omega_k}\int_0^z\frac{dz^\prime}{H(z^\prime)}\right)
\eeq   
and differentiating with respect to the redshift $z$ one finds the relation
\bea
\frac{\partial}{\partial z}\left[(1+z)D_A(z)\right]&=&\frac{c}{H(z)}{\rm{cos}}\left(H_0\sqrt{-\Omega_k}\int_0^z\frac{dz^\prime}{H(z^\prime)}\right)\nonumber\\
&=&D_H(z)\sqrt{1+\frac{\Omega_k H_0^2}{c^2}(1+z)^2D_A(z)^2}.
\eea
This implies that $D_H(z)$ is determined entirely in terms of $D_A(z)$ and its derivative $D^\prime_A(z)$ and the curvature constant $\Omega_k H_0^2/c^2$
\beq
D_H(z)=\frac{D_A(z)+(1+z)d^\prime_A(z)}{\sqrt{1+\frac{\Omega_k H_0^2}{c^2}(1+z)^2d_A^2}}.
\eeq

In terms of the radial and angular BAO scale this relation is
\beq
\Delta z=\frac{\sqrt{1+\frac{\Omega_k}{P^2} (1+z)^2/\Delta\theta^2}}{(1+z)\partial_z(1/\Delta\theta)}.
\eeq
Therefore the radial BAO scale $\Delta$z is entirely determined by the angular BAO scale $\Delta\theta$, its derivative and the constant $\Omega_k/P^2$, which is independent of the redshift $z$.   Therefore while any function $\Delta\theta$ or $\Delta z$ can solve the Friedmann equations for some choice of $\rho(z)$ and $p(z)$, it is not possible to simultaneously arbitrarily fix both.

Similarly, given $\Delta z$ and $\Delta\theta$ at a given redshift, one can solve this expression for the derivative of $\Delta\theta$ and so one can extrapolate the BAO angular scale to slightly higher redshifts.  However it is not possible to extrapolate the radial BAO and so this extrapolation rapidly becomes unreliable.  

This means that the results presented above are independent of any BAO scale measurements at redshifts significantly above that of the Lyman $\alpha$ forest measurement.  Any measurements at higher redshift can be made compatible with any cosmological model explaining the measurements here by choosing $H(z)$ correctly at redshifts between the two sets of measurements, assuming that one places no constraints on $\rho(z)$ and $p(z)$.


\subsection{Cosmic Microwave Background: Strategy}

Following Ref.~\cite{aubourg}, there is a kind of measurement of $D_A$ at very high redshift.  This is given by the scale of the acoustic peak in the cosmic microwave background (CMB) power spectrum.  In the standard cosmological model, this corresponds to the same acoustic oscillation as the BAO peak observed in the matter power spectrum.  However it decoupled not at the drag epoch, but rather at recombination.  As these epochs were quite close in time, there was little movement of the acoustic peak between the two and so one expects the sizes of the features to be similar.  A dramatic relative movement would require extremely early reionization or else a large change in the baryon to photon ratio, which would be in tension, for example, with big bang nucleosynthesis.  Modifications to the cosmological model such as early dark energy or extra neutrinos can therefore shift the relative sizes of the CMB and baryon acoustic peaks by much less than the 7\% anomaly which we are studying here.

In this section we will examine the implications of considering the CMB acoustic peak measurement as a measurement of the BAO feature.  We consider the analysis of the 2015 Planck data in Ref.~\cite{wang15} which finds
\beq
\Delta\theta=0.010392\pm 3\times 10^{-6}.
\eeq  
We include this measurement in our dataset as if it were a measurement of the BAO at the drag epoch redshift, which Planck found to be $z=1059$ in Ref.~\cite{planck15}.   

Critically we avoid placing any uncertainty on $z$, which would have the effect of choosing a level of certainty in our cosmological model and so would ruin the model independent character of our study.  However the price for this assumption is that the uncertainties that we will find for cosmological parameters will be greatly underestimated.  To obtain reliable estimates on cosmological parameters one would need to assign an uncertainty to the relation between the BAO and CMB acoustic scales, which would require strong cosmological assumptions.  According to Ref.~\cite{aubourg}, for the models that they have considered, this additional error is less than $0.1\%$ and so is irrelevant for many purposes, but not all.  Indeed, using this Planck data point and ignoring this uncertainty we naively rule out the best fit Planck $\Lambda$CDM model because Planck is sensitive enough to distinguish the baryonic acoustic peak from the CMB acoustic peak, and so to falsify the crude approximation which underlies this analysis.  

Although the parameter constraints found in this section will therefore be incorrect, they will nonetheless serve to illustrate the complementarity between the BAO and CMB data.  Note that as this is an angular BAO measurement at a much higher redshift than the others, it can be made consistent with any cosmological model designed to fit the BAO data points in the previous sections by correctly choosing $\rho(z)$ and $p(z)$ at $2.34<z<1059$.  Therefore the analysis in this section is not capable of contradicting that in previous sections, unless one makes arbitrary assumptions regarding the functional dependence of the dark energy evolution continuing all of the way back to the drag epoch.

\subsection{Cosmic Microwave Background: Results}

Now including the CMB measurement of the acoustic scale, we will repeat our model testing and parameter estimation.  The results of the model testing with the high $z$ and all data are reported in Tables~\ref{zaltocmbtab} and \ref{tutticmbtab} respectively.  Despite the additional effective degree of freedom, in most cases the models are now in more tension with the data.  In fact, every model except for o$w$CDM is now excluded at more than $2.3\sigma$ by the full data set and $2.8\sigma$ by the high $z$ data set.

\begin{table}
\centering
\begin{tabular}{|c|l|l|l|}
\hline
Model&$\chi^2$&Eff. DOF&$p$-value\\
\hline\hline
$\Lambda$CDM&$14.1$&$3$&$0.0029$\\
\hline
o$\Lambda$CDM&$11.0$&$2$&$0.0040$\\
\hline
$w$CDM&$13.7$&$2$&$0.011$\\
\hline
o$w$CDM&$3.7$&$1$&$0.16$\\
\hline
\end{tabular}
\caption{$p$ values of $z>0.43$ data including the CMB sound horizon measurement with respect to various models.  The effective number of degrees of freedom is the number of data points, 5, minus the number of parameters in the model. The $\Lambda$CDM model is inconsistent with the data at nearly 3$\sigma$ with or without the $\Lambda$CDM-Planck values of the $P$ and $\Omega_m$.  $w$CDM is inconsistent at more than 3$\sigma$.}
\label{zaltocmbtab}
\end{table}

\begin{table}
\centering
\begin{tabular}{|c|l|l|l|}
\hline
Model&$\chi^2$&Eff. DOF&$p$-value\\
\hline\hline
$\Lambda$CDM&$16.8$&$7$&$0.018$\\
\hline
$w$CDM&$15.9$&$6$&$0.014$\\
\hline
o$\Lambda$CDM&$15.0$&$6$&$0.020$\\
\hline
o$w$CDM&$7.5$&$5$&$0.28$\\
\hline
CPL&$15.0$&$5$&$0.010$\\
\hline
\end{tabular}
\caption{$p$ values of all BAO data including the CMB sound horizon measurement with respect to various models.  The effective number of degrees of freedom is the number of data points, 9, minus the number of parameters in the model.}  
\label{tutticmbtab}
\end{table}

\begin{figure} 
\begin{center}
\includegraphics[angle=-0,width=2.8in]{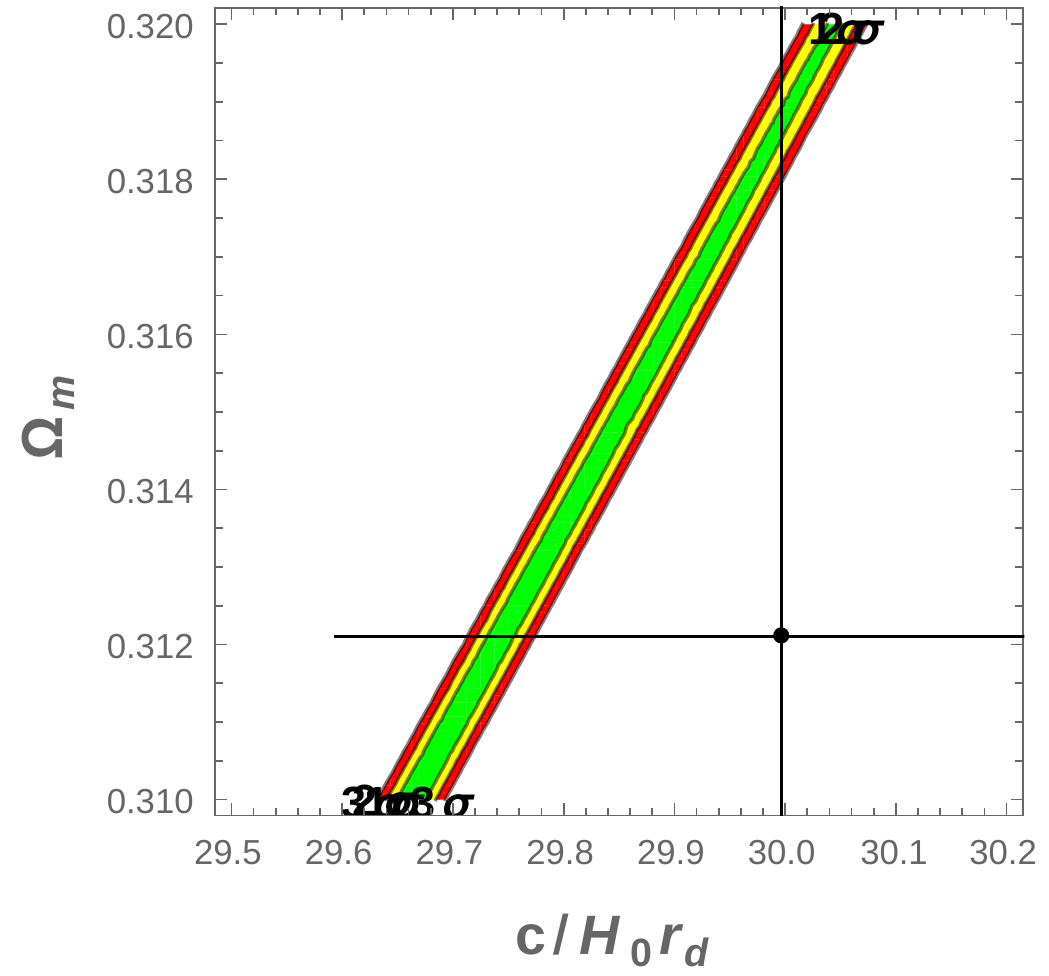}
\includegraphics[angle=-0,width=2.8in]{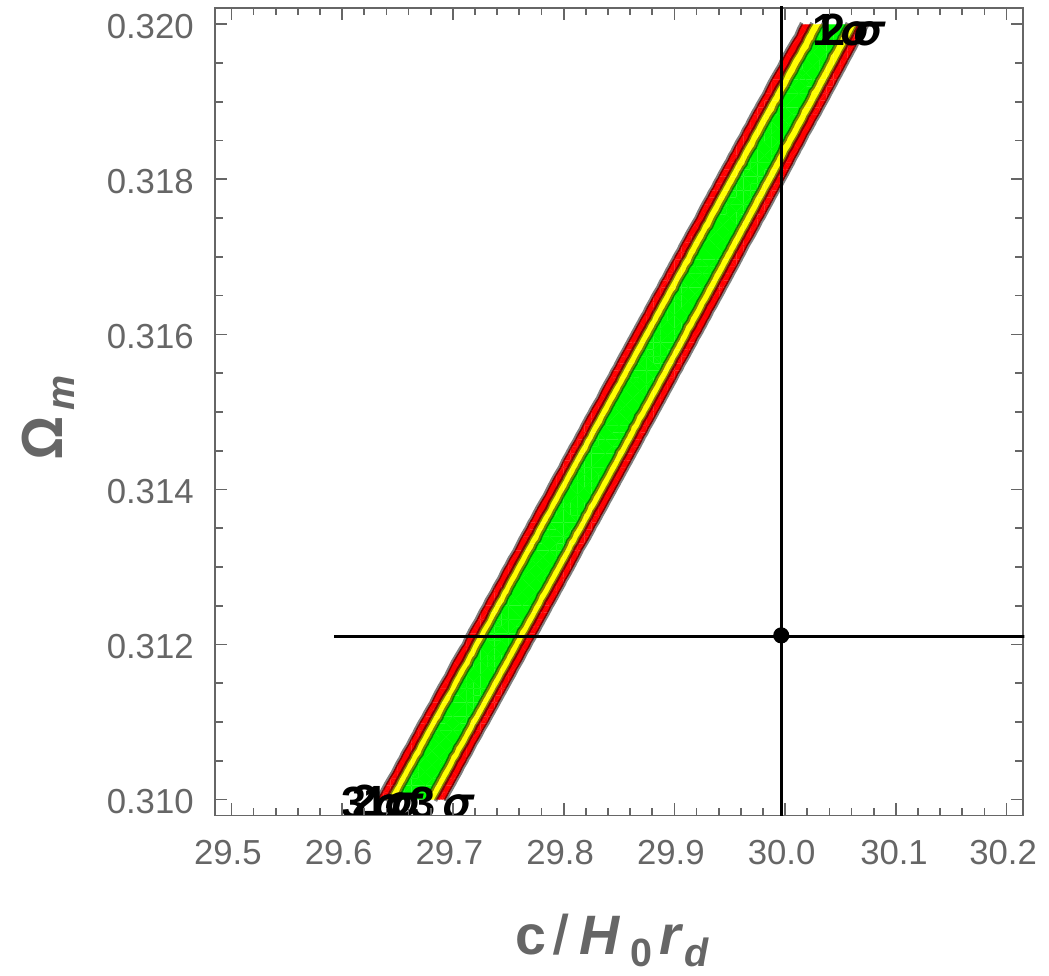}
\caption{The $1\sigma,\ 2\sigma$ and $3\sigma$ confidence regions for a fit of $P=c/(H_0r_d)$ and $\Omega_m$ to the $z>0.43$ data including the CMB sound horizon measurement (left) and to all BAO data including the CMB sound horizon measurement (right), assuming the $\Lambda$CDM model.  The dot is the $\Lambda$CDM-Planck best fit value with its associated uncertainties.}   
\label{lcdmcmbfig}
\end{center}
\end{figure}

\begin{figure} 
\begin{center}
\includegraphics[angle=-0,width=2.8in]{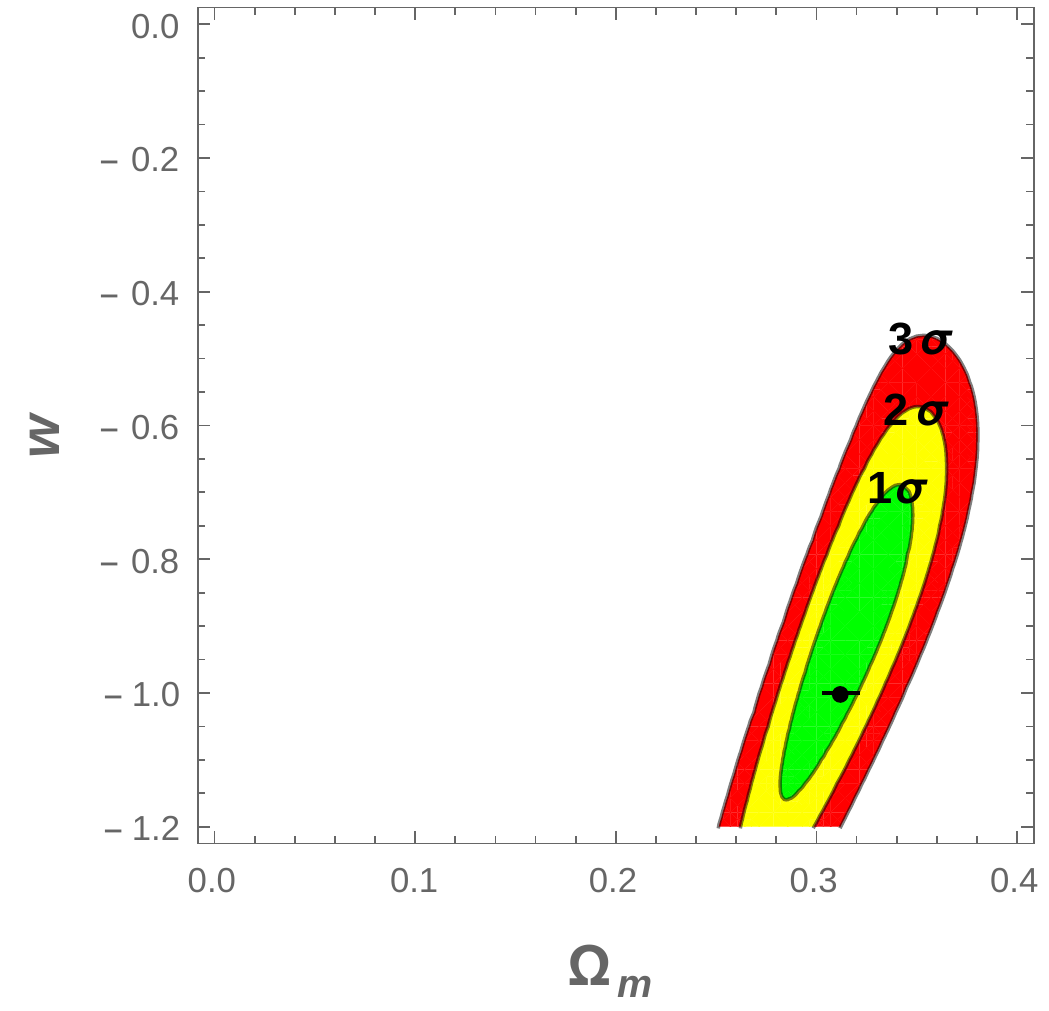}
\includegraphics[angle=-0,width=2.8in]{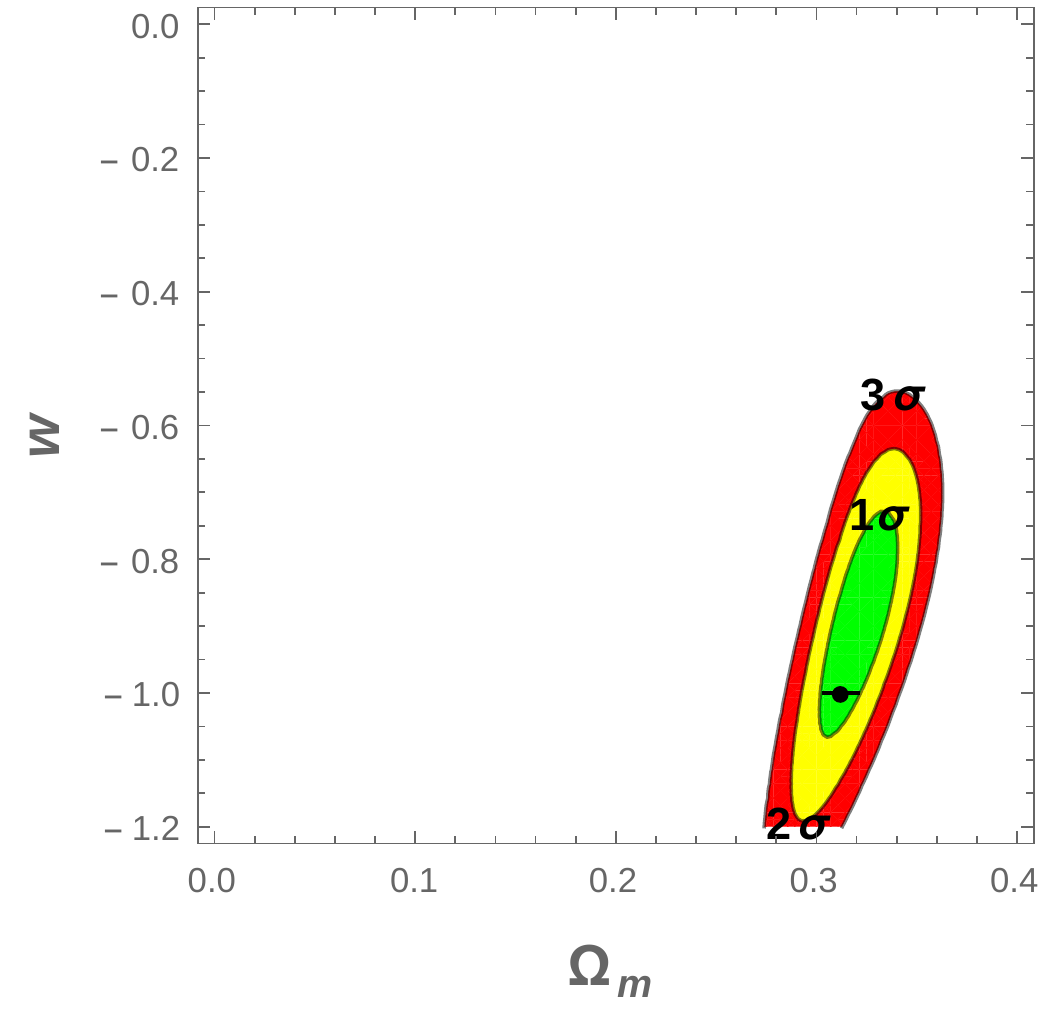}
\caption{The $1\sigma,\ 2\sigma$ and $3\sigma$ confidence regions for a fit of $w$ and $\Omega_m$ to the $z>0.43$ data including the CMB sound horizon measurement (left) and to all BAO data including the CMB sound horizon measurement (right), assuming the $w$CDM model.  The dot is the $\Lambda$CDM-Planck best fit value with its associated uncertainties. }
\label{wcmbfig}
\end{center}
\end{figure}

\begin{figure} 
\begin{center}
\includegraphics[angle=-0,width=2.8in]{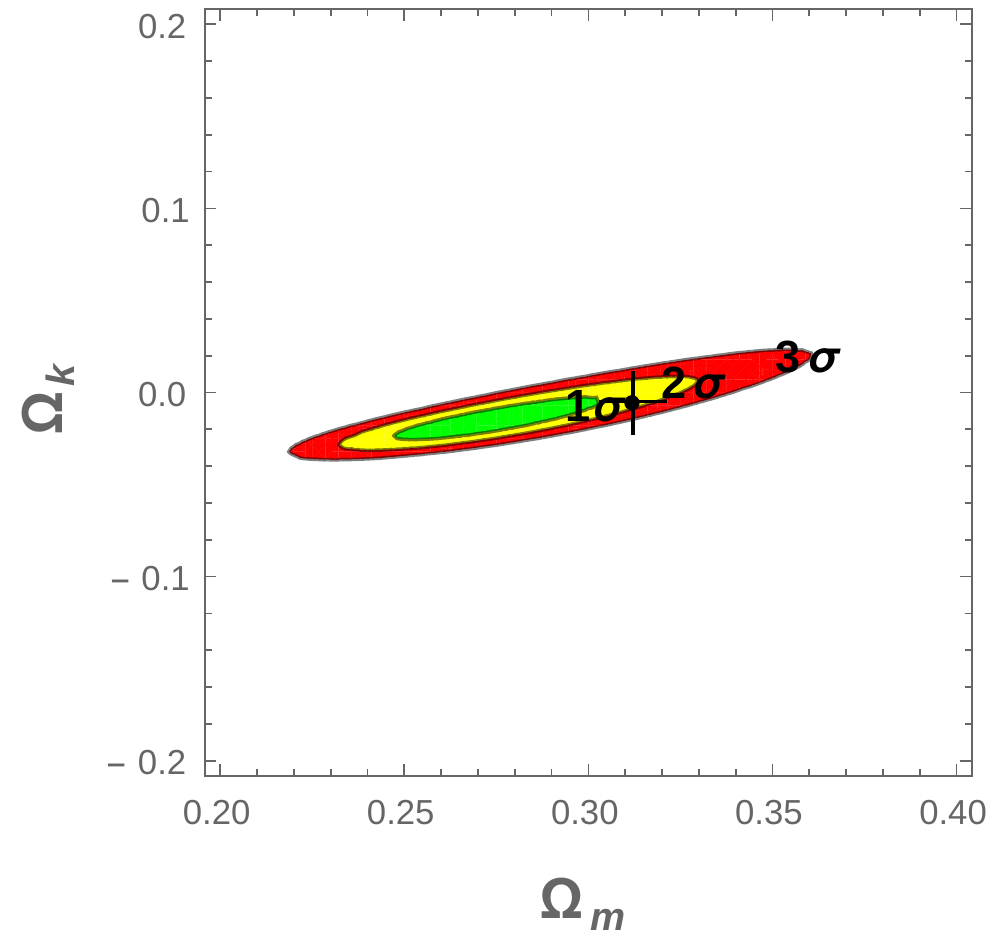}
\includegraphics[angle=-0,width=2.8in]{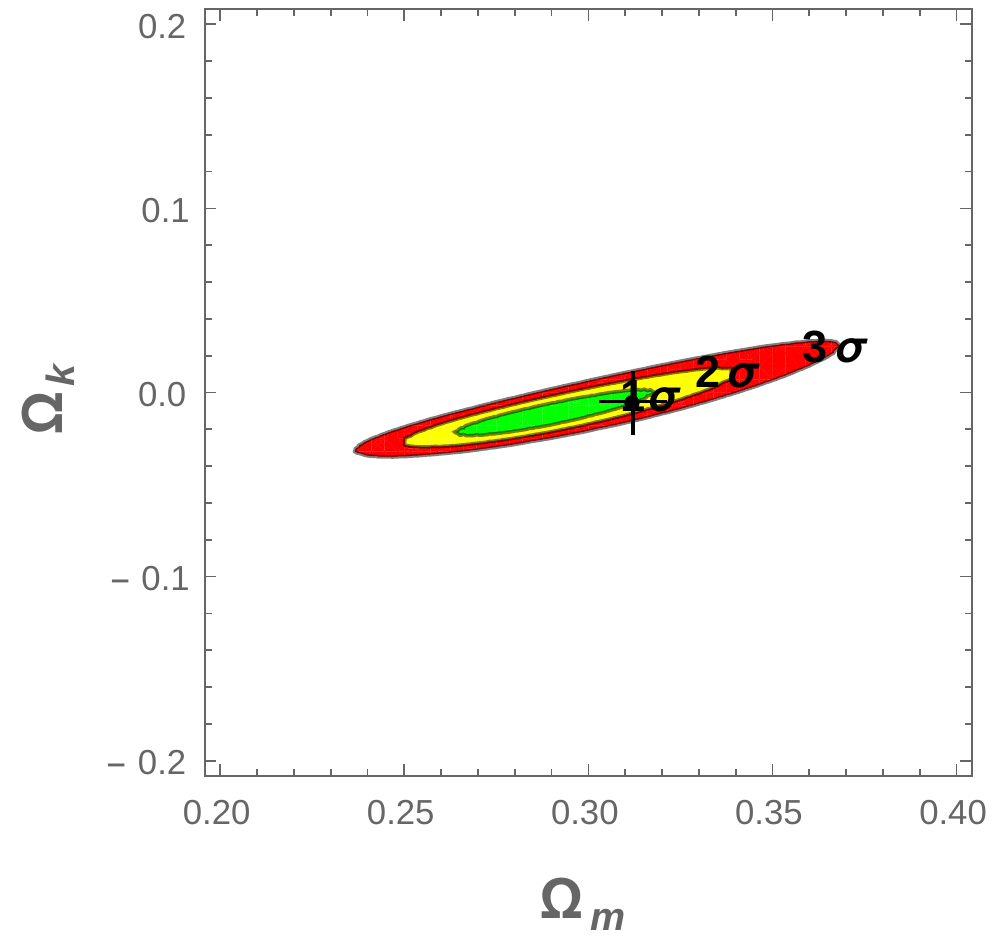}
\caption{The $1\sigma,\ 2\sigma$ and $3\sigma$ confidence regions for a fit of $\Omega_m$ and $\Omega_k$ to the $z>0.43$ data including the CMB sound horizon measurement (left) and to all BAO data including the CMB sound horizon measurement (right), assuming the o$\Lambda$CDM model.  The dot is the $\Lambda$CDM-Planck best fit value with its associated uncertainty shown for $\Omega_m$ and the o$\Lambda$CDM uncertainty, as reported by Planck \cite{planck15}, shown for $\Omega_k$.}   
\label{kcmbfig}
\end{center}
\end{figure}

\begin{figure} 
\begin{center}
\includegraphics[angle=-0,width=2.8in]{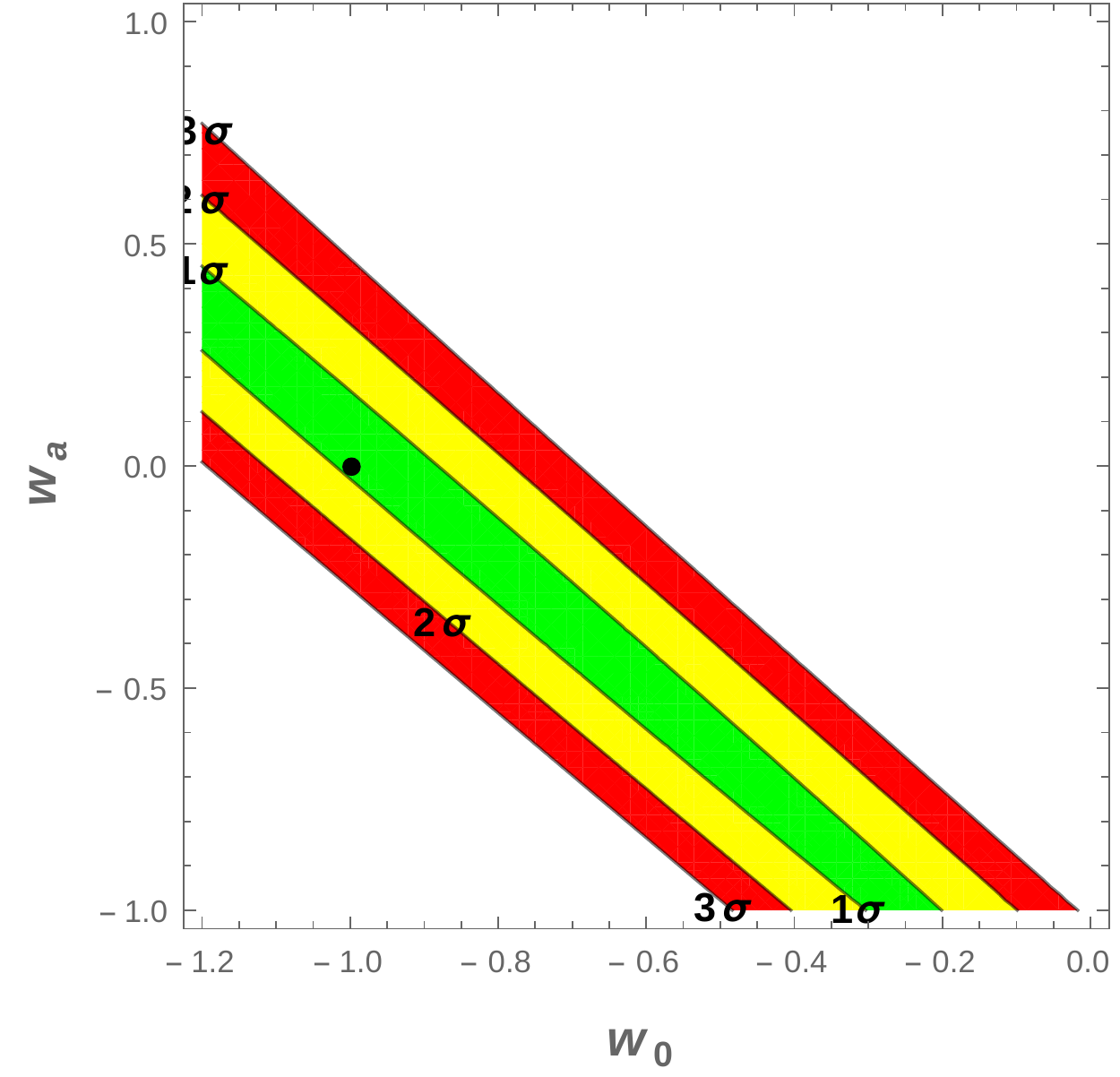}
\caption{The $1\sigma,\ 2\sigma$ and $3\sigma$ confidence regions for a fit of $w_0$ and $w_a$ to all of the BAO data including the CMB sound horizon measurement, assuming the CPL model. }
\label{cplcmbfig}
\end{center}
\end{figure}

The reason for this increased tension is apparent in the new parameter estimations in Figs.~\ref{lcdmcmbfig}, \ref{wcmbfig}, \ref{kcmbfig} and \ref{cplcmbfig}.  The unusual optimal parameter choices which reduced the tension without the CMB data, in particular in the wCDM model, are strongly excluded by the Planck acoustic oscillation measurement.  Once these distant parameter choices are excluded by the CMB data, the models perform poorly.   The spatial curvature is even more strongly constrained, but this has little effect on the significance tests because the BAO data did not prefer a large spatial curvature. 

This in no way affects the evidence for dynamical dark energy presented in the previous sections.  After all, we have argued that they can always be consistent with the CMB sound horizon measurement for some behaviour of $H(z)$ at $2.34<z<1059$.  It simply means that the parametrizations of the equation of state considered there cannot be naively extrapolated back to the damping epoch as in $w$CDM and CPL.  As there was no theoretical motivation for these naive parametrizations, this is not so surprising.

\begin{figure} 
\begin{center}
\includegraphics[angle=-0,width=2.8in]{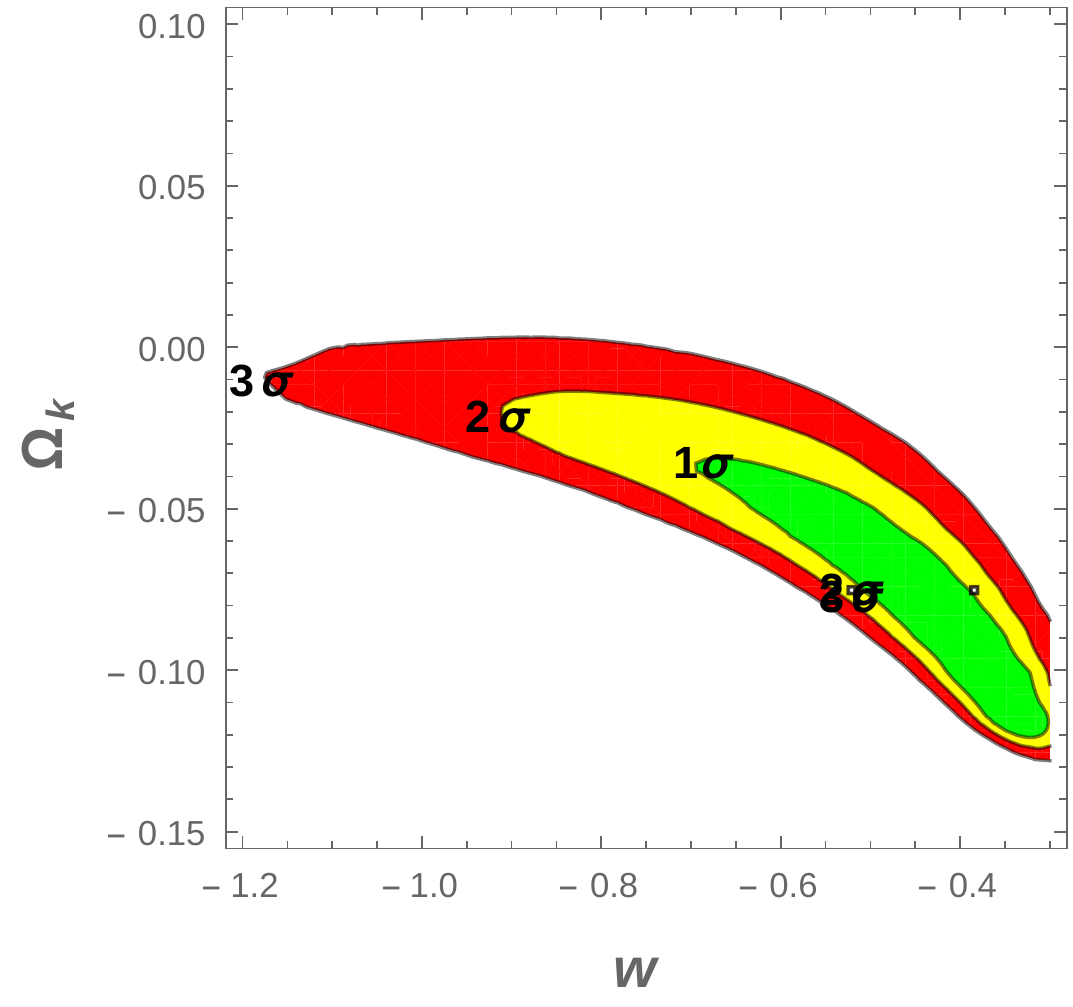}
\includegraphics[angle=-0,width=2.8in]{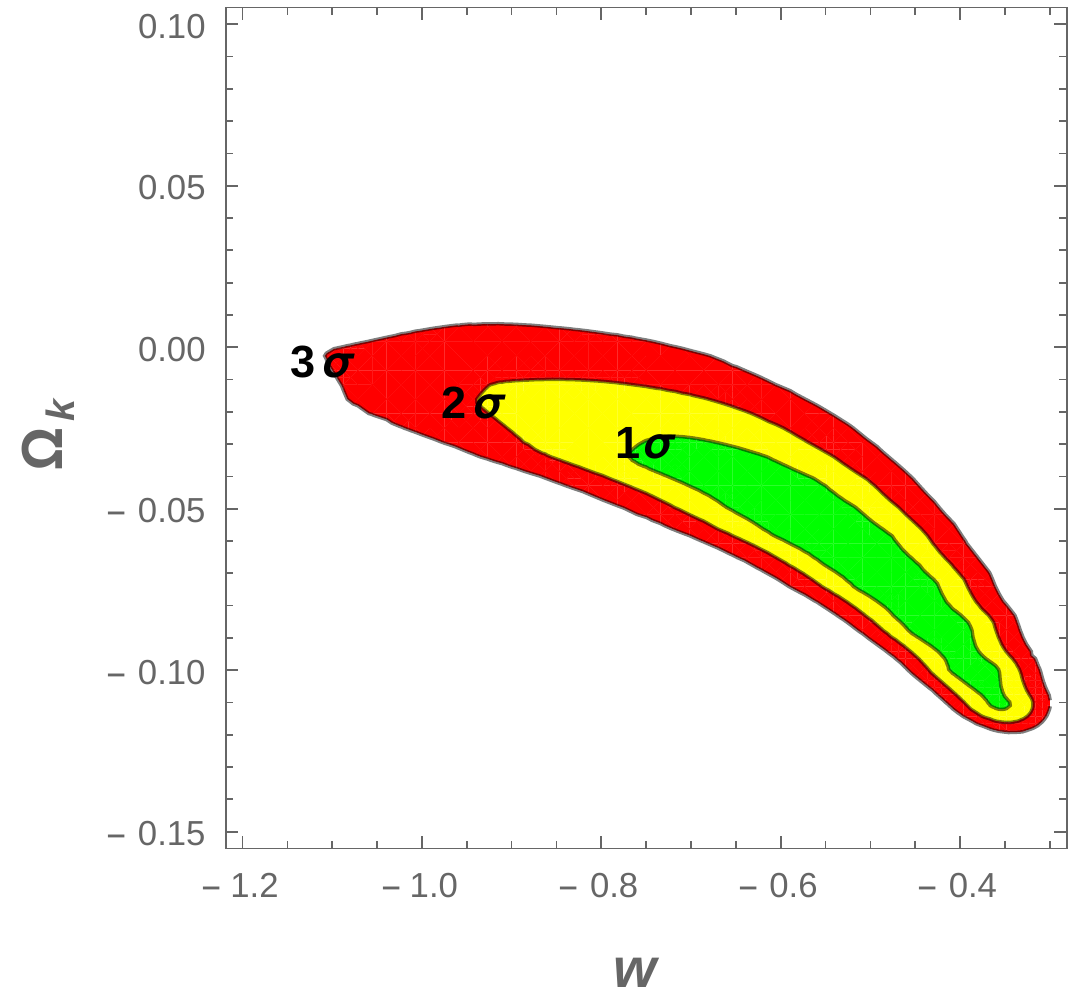}
\caption{The $1\sigma,\ 2\sigma$ and $3\sigma$ confidence regions for a fit of $w$ and $\Omega_k$ to the $z>0.43$ data including the CMB sound horizon measurement (left) and to all BAO data including the CMB sound horizon measurement (right), assuming the o$w$CDM model.}
\label{wkcmbfig}
\end{center}
\end{figure}

The o$w$CDM model has enough free parameters to evade this new constraint.  It manages to provide an acceptable fit by choosing a curvature and dark energy equation of state whose effects somewhat cancel at recombination, but are quite far from the $\Lambda$CDM values, as can be seen in Fig,~\ref{wkcmbfig}.  

\section{Remarks}

An unanchored analysis of BAO is one in which it is assumed that the BAO scale is a standard ruler, but the ruler is not calibrated using any cosmological model.  The unanchored BAO scale is independent of the cosmology at higher redshifts than the observation itself.

In this note we investigated the reported anomaly between the BAO scale measured in the Lyman $\alpha$ forest at an effective redshift of $z=2.34$ and the $\Lambda$CDM-Planck cosmology.  Our goal was to answer the following question: If this anomaly is confirmed by future observations, what must change in the standard cosmological model?

We found that the anomaly is present already in unanchored BAO, implying that this tension is caused by cosmology at $z<2.34$.  In this regime, only the expansion of the Universe, which is characterized by the parameters $P$, $\Omega_m$ and $\Omega_k$ and the function $w(z)$, may affect BAO measurements.  Fixing $P$, $\Omega_m$ and $\Omega_k$ had little effect on the significance of the anomaly, and fits of $\Omega_m$, $\Omega_k$ and $P$ to BAO data resulted in values close to the $\Lambda$CDM-Planck values.  Therefore we concluded that the tension arises from the dark energy equation of state $w(z)$.  However we noticed that its significance decreases from about 3$\sigma$ to 2$\sigma$ when including data at low $z$.  This implied our main result, that the tension is caused by evolution of $w(z)$ in the regime $0.57<z<2.34$.  Indeed the CMB data strongly disfavor a solution with a redshift-independent dynamical dark energy equation of state.

There have been several other analyses of these BAO observations, with an eye on implications for the dark energy equation of state $w(z)$.  In Ref.~\cite{jaber}, the authors examine the same data, including WiggleZ, but consider only the direction-averaged BAO scale $D_V(z)$.  They find much better fits to cosmological models than were found here using the full two-dimensional constraints on the $D_A(z)-D_H(z)$ plane.  The reason for the discrepancy is that, as explained in Ref.~\cite{ffbao}, the Lyman $\alpha$ anomaly arises from a lower than expected observed value of $D_A(z)/r_d$ and a higher value of $D_H(z)/r_d$.  However in the isotropic $D_V(z)$ these two anomalies somewhat cancel one another.

Ref.~\cite{mebao} used the BOSS data but only considered $D_H(z)/r_d$.  This approach had the advantage that no model for the evolution of $w(z)$ needed to be assumed in the analysis, since, unlike $D_A(z)$, $D_H(z)$ can be expressed directly in terms of an average equation of state \cite{wmap5}.  As a result, it was shown that the tension can be reduced with an equation of state which starts high at low redshift but becomes more negative at higher redshift, a claim which was shown here only in the special case of the CPL parametrization.

However, $D_H(z)/r_d$ is only a single number at each redshift.  Therefore the high redshift sample, where the anomaly is most evident, consisted of only two data points.  Using these two data points, one determined the over all scale $P$, and so only a single data point, the ratio, constrains both $w(z)$ and $\Omega_m$.  The resulting degeneracy between $w(z)$ and $\Omega_m$ meant that the study \cite{mebao} needed to rely on an assumed regime for $\Omega_m$ to robustly confirm the existence of the anomaly.  In all, the $D_H(z)$-only analysis provided a complimentary approach to that presented in the present paper, as the confirmation of the anomaly was less robust but the implications for the behavior of $w(z)$ were independent of any assumed parametrization.

\section* {Acknowledgement}
\noindent
JE is supported by NSFC MianShang grant 11375201.



\begin{thebibliography}{99}


\bibitem{baoteor}
 H.~J.~Seo and D.~J.~Eisenstein,
  ``Probing dark energy with baryonic acoustic oscillations from future large galaxy redshift surveys,''
  Astrophys.\ J.\  {\bf 598} (2003) 720
  [astro-ph/0307460].

\bibitem{eisenrecon}
    D.~J.~Eisenstein, H.~J.~Seo, E.~Sirko and D.~Spergel,
  ``Improving Cosmological Distance Measurements by Reconstruction of the Baryon Acoustic Peak,''
  Astrophys.\ J.\  {\bf 664} (2007) 675
  [astro-ph/0604362].

\bibitem{recon2}
   N.~Padmanabhan and M.~White,
  ``Calibrating the Baryon Oscillation Ruler for Matter and Halos,''
  Phys.\ Rev.\ D {\bf 80} (2009) 063508
  [arXiv:0906.1198 [astro-ph.CO]].


\bibitem{ffbao}
 T.~Delubac {\it et al.} [BOSS Collaboration],
  ``Baryon acoustic oscillations in the Lyα forest of BOSS DR11 quasars,''
  Astron.\ Astrophys.\  {\bf 574} (2015) A59
  [arXiv:1404.1801 [astro-ph.CO]].

\bibitem{fqbao}
  A.~Font-Ribera {\it et al.} [BOSS Collaboration],
  ``Quasar-Lyman $\alpha$ Forest Cross-Correlation from BOSS DR11 : Baryon Acoustic Oscillations,''
  JCAP {\bf 1405} (2014) 027
  [arXiv:1311.1767 [astro-ph.CO]].

\bibitem{gongboten}
  G.~B.~Zhao {\it et al.},
  ``The clustering of galaxies in the completed SDSS-III Baryon Oscillation Spectroscopic Survey: Examining the observational evidence for dynamical dark energy,''
  arXiv:1701.08165 [astro-ph.CO].

\bibitem{aubourg}
  E.~Aubourg {\it et al.},
  ``Cosmological implications of baryon acoustic oscillation measurements,''
  Phys.\ Rev.\ D {\bf 92} (2015) no.12,  123516
  [arXiv:1411.1074 [astro-ph.CO]].

\bibitem{addison}
  G.~E.~Addison, G.~Hinshaw and M.~Halpern,
  ``Cosmological constraints from baryon acoustic oscillations and clustering of large-scale structure,''
  Mon.\ Not.\ Roy.\ Astron.\ Soc.\  {\bf 436} (2013) 1674
  [arXiv:1304.6984 [astro-ph.CO]].

\bibitem{mebao}
 J.~Evslin,
  ``Model-Independent Dark Energy Equation of State from Unanchored Baryon Acoustic Oscillations,''
  Phys.\ Dark Univ.\  {\bf 13} (2016) 126
  [arXiv:1510.05630 [astro-ph.CO]].

\bibitem{riess}
  A.~G.~Riess {\it et al.},
  ``A 2.4\% Determination of the Local Value of the Hubble Constant,''
  Astrophys.\ J.\  {\bf 826} (2016) no.1,  56
  [arXiv:1604.01424 [astro-ph.CO]].
  

\bibitem{6df}
   F.~Beutler {\it et al.},
  ``The 6dF Galaxy Survey: Baryon Acoustic Oscillations and the Local Hubble Constant,''
  Mon.\ Not.\ Roy.\ Astron.\ Soc.\  {\bf 416} (2011) 3017
  [arXiv:1106.3366 [astro-ph.CO]].

\bibitem{dr7}
    A.~J.~Ross, L.~Samushia, C.~Howlett, W.~J.~Percival, A.~Burden and M.~Manera, ``The clustering of the SDSS DR7 main Galaxy sample – I. A 4 per cent distance measure at $z = 0.15$,''
  Mon.\ Not.\ Roy.\ Astron.\ Soc.\  {\bf 449} (2015) no.1,  835
  [arXiv:1409.3242 [astro-ph.CO]].

\bibitem{wigglez}
   C.~Blake {\it et al.},
  ``The WiggleZ Dark Energy Survey: mapping the distance-redshift relation with baryon acoustic oscillations,''
  Mon.\ Not.\ Roy.\ Astron.\ Soc.\  {\bf 418} (2011) 1707
  [arXiv:1108.2635 [astro-ph.CO]].

 
\bibitem{bosspower}
H.~Gil-Marín {\it et al.},
  ``The clustering of galaxies in the SDSS-III Baryon Oscillation Spectroscopic Survey: BAO measurement from the LOS-dependent power spectrum of DR12 BOSS galaxies,''
arXiv:1509.06373 [astro-ph.CO].

\bibitem{bosscorr}
A.~J.~Cuesta {\it et al.},
  ``The clustering of galaxies in the SDSS-III Baryon Oscillation Spectroscopic Survey: Baryon Acoustic Oscillations in the correlation function of LOWZ and CMASS galaxies in Data Release 12,''
  Mon.\ Not.\ Roy.\ Astron.\ Soc.\  {\bf 457} (2016) 1770
  [arXiv:1509.06371 [astro-ph.CO]].

\bibitem{chitab}
http://darkmatter.ps.uci.edu/baofit/

\bibitem{ly13}
 J.~E.~Bautista {\it et al.},
  ``Measurement of BAO correlations at $z=2.3$ with SDSS DR12 Ly$\alpha$-Forests,''
  arXiv:1702.00176 [astro-ph.CO].
  
\bibitem{planck15}
   P.~A.~R.~Ade {\it et al.} [Planck Collaboration],
  ``Planck 2015 results. XIII. Cosmological parameters,''
   arXiv:1502.01589 [astro-ph.CO].
    
\bibitem{dvnunes}
  R.~C.~Nunes, S.~Pan and E.~N.~Saridakis,
  ``New constraints on interacting dark energy from cosmic chronometers,''
  Phys.\ Rev.\ D {\bf 94} (2016) no.2,  023508
  doi:10.1103/PhysRevD.94.023508
  [arXiv:1605.01712 [astro-ph.CO]].
  
\bibitem{heavens}
  A.~Heavens, R.~Jimenez and L.~Verde,
  ``Standard rulers, candles, and clocks from the low-redshift Universe,''
  Phys.\ Rev.\ Lett.\  {\bf 113} (2014) no.24,  241302
  [arXiv:1409.6217 [astro-ph.CO]].
  
\bibitem{dvmuk}
  A.~Mukherjee and N.~Banerjee,
  ``Parametric reconstruction of the cosmological jerk from diverse observational data sets,''
  Phys.\ Rev.\ D {\bf 93} (2016) no.4,  043002
  [arXiv:1601.05172 [gr-qc]].

\bibitem{dvbas}
  S.~Basilakos,
  ``Linear growth in power law $f(T)$ gravity,''
  Phys.\ Rev.\ D {\bf 93} (2016) no.8,  083007
  [arXiv:1604.00264 [gr-qc]].

\bibitem{covar}
  R.~Emami, D.~Grin, J.~Pradler, A.~Raccanelli and M.~Kamionkowski,
  ``Cosmological tests of an axiverse-inspired quintessence field,''
  Phys.\ Rev.\ D {\bf 93} (2016) no.12,  123005
  [arXiv:1603.04851 [astro-ph.CO]].
 
\bibitem{cpl1}
M.~Chevallier and D.~Polarski,
  ``Accelerating universes with scaling dark matter,''
  Int.\ J.\ Mod.\ Phys.\ D {\bf 10} (2001) 213
  [gr-qc/0009008].
  
\bibitem{cpl2}
E.~V.~Linder,
  ``Exploring the expansion history of the universe,''
  Phys.\ Rev.\ Lett.\  {\bf 90} (2003) 091301
  [astro-ph/0208512].


\bibitem{lee}
   L.~Lyons,
  ``Open statistical issues in Particle Physics,''
  Ann.\ Appl.\ Stat.\  {\bf 2} (2008) 887.

\bibitem{pi}
   A.~Frolop and D.~Scott,
  ``Pi in the sky,''
  arXiv:1603.09703 [astro-ph.CO].
  
\bibitem{jaber}
  M.~Jaber and A.~de la Macorra,
  ``Constraints on Steep Equation of State for the Dark Energy using BAO,''
  arXiv:1604.01442 [astro-ph.CO].
    
\bibitem{1405.5116}
  V.~H.~Cardenas,
  ``Exploring hints for dark energy density evolution in light of recent data,''
  Phys.\ Lett.\ B {\bf 750} (2015) 128
  doi:10.1016/j.physletb.2015.08.064
  [arXiv:1405.5116 [astro-ph.CO]].
  
\bibitem{greci}
   G.~Pantazis, S.~Nesseris and L.~Perivolaropoulos,
  ``A Comparison of Thawing and Freezing Dark Energy Parametrizations,''
  arXiv:1603.02164 [astro-ph.CO].
  
\bibitem{gongboeboss}
  G.~B.~Zhao {\it et al.},
  ``The extended Baryon Oscillation Spectroscopic Survey (eBOSS): a cosmological forecast,''
  Mon.\ Not.\ Roy.\ Astron.\ Soc.\  {\bf 457} (2016) 2377
  [arXiv:1510.08216 [astro-ph.CO]].

\bibitem{wang15}
Y.~Wang and M.~Dai,
  ``Exploring uncertainties in dark energy constraints using current observational data with Planck 2015 distance priors,''
  Phys.\ Rev.\ D {\bf 94} (2016) no.8,  083521
  [arXiv:1509.02198 [astro-ph.CO]].
 
\bibitem{wmap5}
  E.~Komatsu {\it et al.} [WMAP Collaboration],
  ``Five-Year Wilkinson Microwave Anisotropy Probe (WMAP) Observations: Cosmological Interpretation,''
  Astrophys.\ J.\ Suppl.\  {\bf 180} (2009) 330
  [arXiv:0803.0547 [astro-ph]].


  
\end{thebibliography}
\end{document}